\documentclass[journal]{IEEEtran}

\ifCLASSINFOpdf

\else

\fi

\usepackage{amssymb}
\usepackage{graphicx}
\usepackage{subfigure}
\usepackage{float}
\usepackage{amsmath}
\usepackage{booktabs}
\usepackage[numbers,sort&compress]{natbib}
\usepackage{array}
\newcolumntype{P}[1]{>{\centering\arraybackslash}p{#1}}
\begin{document}
\title{Exploiting AIS Data for Intelligent Maritime Navigation: A Comprehensive Survey}

\author{Enmei Tu,
        Guanghao Zhang,
        Lily Rachmawati,
        Eshan Rajabally,
        Guang-Bin Huang,~\IEEEmembership{Senior Member,~IEEE}
\thanks{Enmei Tu is with the Rolls-Royce@NTU Corporate Lab, Nanyang Technological University, Singapore}
\thanks{Guanghao Zhang and Guang-Bin Huang are with School of Electrical \& Electronic Engineering, Nanyang Technological University, Singapore}
\thanks{Lily Rachmawati is with Computational Engineering Team, Advanced Technology Centre,  Rolls-Royce Singapore Pte Ltd}
\thanks{Eshan Rajabally is with Strategic Research Center, Rolls-Royce Plc}}
\markboth{}%
{Tu \MakeLowercase{\textit{et al.}}: Survey of Maritime Autonomous Navigation}

\maketitle

\begin{abstract}
The Automatic Identification System (AIS) tracks vessel movement by means of electronic exchange of navigation data between vessels, with onboard transceiver, terrestrial and/or satellite base stations. The gathered data contains a wealth of information useful for maritime safety, security and efficiency. This paper surveys AIS data sources and relevant aspects of navigation in which such data is or could be exploited for safety of seafaring, namely traffic anomaly detection, route estimation, collision prediction and path planning.
\end{abstract}

\IEEEpeerreviewmaketitle

\section{Introduction}
\IEEEPARstart{M}aritime transportation provides the most energy efficient means of transporting large quantities of goods over large distances. The central role of maritime transportation in the world's logistic system is evident in the statistical estimates from International Maritime Organization (IMO): around 90\% of world trade is carried by sea and the trade volume are still growing at a rate even faster than global economy \cite{kaluza2010complex}. Growth in world economy and trading translates into increasing demand for more ships with larger cargo capacity and higher travelling speed and highlights concerns in maritime safety and security.

The rapid increase in the affordability of data acquisition, storage and processing infrastructure and advances in intelligent techniques to learn from data offer means to significantly improve maritime safety and security while reducing costs \cite{weintrit2009marine}.

At sea, conventional maritime navigation relies on human judgement by the officer on watch assisted by a chart plotter, radar, sonar and the occasional closed-circuit television (CCTV) and/or infra red cameras. The reliance on manual surveillance, interpretation and decision making means that crew fatigue from overwork and lack of experience and/or skills in junior hires can significantly compromise safety. Statistical study shows that 75\%-96\% of marine accidents are caused by human errors \cite{harati2007automatic, lazarowska2012decision}. While engaging seasoned crew will reduce the rate of accidents, crewing costs, currently accounting for approximately 35\%-68\% of the daily operating cost\footnote{Operating costs usually include crew, stores and lubes, maintenance and repair, insurance costs and overhead costs and are often distinguished from voyage costs such as fuel and bunkering cost.} \cite{ComparisonMarad}, will increase in the future as seafaring jobs become comparatively less attractive.

In ports, human officers monitor traffic patterns from various surveillance means, identify suspicious patterns or potential collisions and raise alerts. Advanced ports, like the Port of Singapore, utilize Vessel Traffic Management System to gather real time vessel information - fusing vessel database, navigational data and electro optical camera surveillance. Trained officers then monitor the displayed information for safety and security risks to pre empt suspect vessels and/or activities.

Intelligent analyses of Automatic Identification System (AIS) can potentially enhance and/or replace manual lookout and surveillance both at sea and in port.
AIS recently has been made compulsory for marine safety for international commercial ships above a certain tonnage, including cargo, passage, tankers, etc. Integrating a standardized Very High Frequencies (VHF) transceiver, AIS broadcasts kinematic information (including ship location, speed, course, heading, rate of turn, destination and estimated arrival time) as well as static information (including ship name, ship MMSI ID, message ID, ship type, ship size, current time) of its host ship every 2 to 10 seconds, depending on the host ship's speed while underway, and every 3 minutes while the host ship is at anchor \cite{tetreault2005use}. Meanwhile, the transceiver continuously collects these information broadcasted from other ships within 20 nautical miles range to its host ship. These information can also be transferred in long distance by means of terrestrial and/or satellite base stations.

Real-time and historical AIS data contains potentially useful markers for the early identification of anomalous activities or vessels and collision risk. There are also challenges in extracting knowledge from AIS data arising from the volume of the data, incompleteness, noise, rogue/dark vessels etc \cite{last2014comprehensive}. Significant work has been done in the maritime intelligent technology community on extracting valuable information from AIS data. This paper surveys AIS data sources and relevant aspects of navigation in which such data is or could be exploited for safety of seafaring\footnote{It should be mentioned that there are also other types of data (such as radar, video etc.) that can be used for these applications, but their corresponding algorithms and mechanisms are quite different from that of AIS based and thus are out of the scope of this paper.}, namely traffic anomaly detection, route estimation, collision prediction and path planning. More specifically,
\begin{itemize}
  \item Real time anomaly detection can identify potential security and navigation safety hazards and therefore is valuable for an onboard intelligent navigation system and for port authorities. Anomaly detection aims to mine typical motion patterns from historical data and identify suspicious ship(s) that deviate significantly from the common motion patterns. While normal ship motion is largely predictable as it generally follows a pattern, the irregular motion characteristics of anomalous ships are less predictable. Such vessels present higher collision risks particularly in high traffic density areas, such as busy ports and traffic lanes.
  \item  Route estimation involves constructing a model of vessel motion from historical data and predicting its future trajectory. While short term route estimation is largely achievable, medium term and longer term estimation are more useful given the restricted maneuverability of some types of vessels, e.g. container vessels and bulk carriers.
  \item Collision prediction assesses the collision risk between own ship and other target ships based on the predicted trajectories. If two trajectories have an intercross, the collision risk is very large and a collision may happen.
  \item  If the collision risk is beyond a certain threshold,  path planning component plans an alternative safe route with minimal cost regarding to the sailing time and distance for own ship to avoid potential collision.
\end{itemize}
In addition, we also present a brief survey of various AIS-data providers/sources on internet accompanied with an initial assessment of their corresponding data quality and availability.



The remainder of the paper is structured as follows. Section 2 describes various AIS data types and characteristics as well as data sources. Sections 3 and 4 survey anomaly detection and route estimation from AIS data. Section 5 and 6 presents collision risk assessment from AIS data and ship path planning, respectively, followed by discussions and conclusions in Section 7. In each case, the survey focuses on representative and most recent work, describing key parts of the algorithm involved and highlighting the advantages as well as disadvantages within the application context.

\section{AIS data sources survey}
\begin{table*}[!t]
	\renewcommand{\arraystretch}{1.3}
	\caption{Data quality comparison}
	\label{ComparisonTable}
	\centering
	\begin{tabular}{|c|P{1.1cm}|P{1.8cm}|P{1.5cm}|P{2cm}|P{2cm}|P{1.8cm}|P{1cm}|P{1cm}|}
	\hline
	\textbf{Data Source} & \textbf{Historical or Live} & \textbf{Data Completeness} & 	\textbf{Validity of Heading} &  \textbf{Estimated Time Res.(min.)} & \textbf{Declared Time Res.(min.)} & \textbf{Pos. Precision (deg.)}  & \textbf{Data Type}  & \textbf{Access Method}\\
    \hline \hline
Marine T. & H/L & Complete & Unknown & 5 & 2 & Full  & T/Com & FTP\\
\hline
IHS G. & H & Complete & 100\% & 40 & \begin{tabular}{@{}c@{}}AIS-T: 3 \\ AIS-S: 360\end{tabular} & Full  & T/Com & FTP\\
\hline
exactE. & H & Complete & 100\% &17 & Unknown & Full  & S/Com & FTP \\
\hline
VT E. & H & Complete & 29.4\% &5 & 5 & Full  & T & FTP \\
\hline

FleetMon & H/L & Complete & 100\% & $5/60^{\#}$  & Unknown  &  Full  & T & API \\
\hline \hline
MarineC. & H & Complete & 45.3\%  &1 & 1 &Full & Com & FTP \\
\hline
Sailwx & H & Missing CoG, Heading & 0\% & 111 &  Unknown & 0.001   & Com & WWW \\
\hline
Aishub & L & Complete & 66.5\% & 18 & Unknown & 0.0001-0.001 & T & WWW \\
\hline
Ais E. & L & Missing time stamp & 37.6\% & 0.24 & Unknown & Full  & T & WWW \\
\hline
Aprs. & L & Complete & 100\% & 5 & Unknown & 0.0001 & T & API \\
\hline
\end{tabular}
\end{table*}

\subsection{Overview}
In this section we present a brief survey of the most popular AIS data sources (of commercial and non-commercial) and their corresponding data quality assessments, since low data quality will bring big challenges to intelligent maritime navigation or data mining system.


AIS data can be received by surrounding ships, terrestrial-based AIS stations located along coast and satellite AIS stations. The typical coverage range of a terrestrial AIS station or an onboard transceiver is about 15-20 nautical miles (nm), depending on many factors such as transceiver location/type or weather conditions. In open seas, satellite-based receivers provide an efficient supplement while terrestrial stations are out of range. In this paper, an AIS message received by terrestrial station is called AIS-Terrestrial record (or AIS-T for short) and a message by satellite is called AIS-Satellite record (AIS-S).

Among all the 27 AIS message types (Listed in Table \ref{SummaryofAIStypes} in appendices for reader's convenience.), position report AIS message is usually  more frequently utilized for common vessel because of its importance in both navigational purpose and data mining research. There are many fields in an position report message (See Table \ref{positionreport} for full data fields' description \cite{navcenter} in an AIS position report), and fields like position, time stamp, speed over ground (SoG), and course over ground (CoG) are of particular interest in intelligent navigation applications.

\subsection{Commercial Data Providers}
The discussion in this section covers five most frequently cited commercial providers of historical AIS data: Marine Traffic (Marine T.) \cite{marinetraffic}, VT explorer (VT E.) \cite{vtexplorer}, IHS global (IHS G.) \cite{ihsglobal}, exactEarth (exactE.) \cite{exactEarth} and FleetMon \cite{fleetmon}. From these commercial companies, customer can obtain either historical data by FTP transfer or live data by API service. For live data, customer may request live positions of vessels in a specific region or with particular MMSI (Maritime Mobile Service Identity) continuously for several weeks or months for enough volume of AIS data for training model. Particularly researcher may test their intelligent navigation system by live data to verify efficiency or if overfitting occurs. Marine Traffic \cite{marinetraffic} and FleetMon \cite{fleetmon} offer both historical and live data.

According to our sample data sets, the longitude and latitude fields of the AIS data are with full precision (1/10000 degree) for all commercial data providers, but the resolution of time stamp, which is important for efficient and effective data mining, can only be guaranteed in VT E. data sets.

\subsection{Free Data Sources}
Survey on free AIS data sources covers five popular providers:  Marinecadastre (MarineC.) \cite{marinecadastre}, Sailwx \cite{sailwx}, Aishub \cite{aishub}, Ais Exploratorium (Ais E.) \cite{aisexploratorium} and Aprs \cite{aprs}. Based on our experiences, one most serious problem of free data sources is that data quality (such as completeness of message or time stamp resolution) usually can not be guaranteed. A brief summary of the characteristics of each source is described as follows:

\begin{itemize}
\item MarineC.: The data source contains historical records from 2009 to 2014 in America. Records are filtered to 1 minute and stored in monthly file by Universal Transverse Mercator (UTM) zone. For the sake of privacy, the ship name and call sign fields are removed and MMSI field is encrypted.
\item Sailwx: One can search position track history of a vessel by its call sign or MMSI. Sailwx also provides some more  useful information such as weather, wind, wave and temperature, but the quality of the sample data is very low, e.g. almost all sample data lack CoG and Heading fields.
\item Aishub: One most important feature of this provider is that people can search all vessels records around a given terrestrial AIS station and its website's can provide a timely response query service.
\item Ais E.: This provider has a share center of real-time forwarded raw AIS data (NMEA format) collected by  one AIS base station in Los Angeles. Decoding raw NMEA messages can be performed by AisDecoder \cite{aisdecoder} if using this data source.
\item Aprs: It offers a free API service  for fetching  latest message given specific vessel MMSI or call sign, but the messages is not necessarily realtime. Another main problem is that the time interval between  consecutive messages may vary from several minutes to hours or even months.
\end{itemize}
\subsection{Data Quality Comparison}
Data quality analysis is conducted based on sample data retrieved from each source. The sample data were collected by various methods, such as requesting for commercial providers, writing program to crawling data or searching on their website. The sample data size varies greatly, from hundred of records to millions of records, depending on the way of collection:
\begin{itemize}
\item Marine T. , Marine T.API and Fleet M.API: There is no official sample data. Evaluation comes from randomly selecting ships' historical and live data (hundreds of records) on website directly.
\item IHG., exactE. and VT E.: Providers offer sample data with thousands of records.
\item MarineC.: Millions of AIS records have been downloaded from the database and exported into ASCII file.
\item Sailwx: Ships are searched according to their MMSI or call sign and their position histories (thousands of lines) are manually checked online.
\item Aishub:  A crawler has been coded to collect millions of ais data on internet from three terrestrial AIS stations in China and Netherlands automatically.
\item Ais E.: Thousands of raw NMEA feeds within about 90 minutes have been collected and decoded by open source software  AisDecoder \cite{aisdecoder}.
\item Aprs.:   100 (randomly selected) vessels' last two position records have been requested online.
\end{itemize}

Assessment is mainly concentrated on position precision (longitude and latitude), time stamp resolution (time interval between two consecutive AIS messages) and data completeness. Note that data quality description may be deviated to some extent due to limitation of volume of sample data. Detailed data quality comparison is shown in table \ref{ComparisonTable}. Description for each column is as below:

\begin{itemize}
\item Data Source: abbreviation for each provider.
\item Historical/Live: historical or live data can be obtained from the provider. 'H/L' represents both historical and live data services.
\item Data completeness: completeness of  AIS message fields in the data,  including position, SoG, CoG, Heading and time stamp.
\item Validity of Heading: percentage of invalid heading field (511 indicates not available). 'Unknown' represents heading field is not included in sample data.
\item Estimated Mean Time Resolution: estimated time stamp resolution by sample data. Note that for FleetMon '5/60' is marked because although estimated mean time resolution of historical data is 5 minutes, API service will downsample to 60 minutes.
\item Declared Time Resolution: providers' declaration of time stamp resolution.
\item Position Precision (deg.): 'Full' means full precision as stated in AIS protocol (1/10000 min.). Actually, due to the limitation of positioning system (GPS), position accuracy is already high if less than 10 meters (around 0.0001 degree, changed due to different position). For Aishub, '0.0001-0.001' range is shown as precision differs for different AIS base stations.
\item Data Type: The type of AIS data can be obtained from that  provider.  'T', 'S' mean AIS-Terrestrial data and AIS-Satellite data, respectively.  'Com' means that a vessel trajectory may be a combination of AIS-T data and AIS-S data.
\item Access Method: 'FTP', 'API', and 'WWW' represent fetching data by FTP transfer, open API service and browsing on website respectively. 'WWW' methods may require crawler program to fetch data automatically.
\end{itemize}

\section{Anomaly Detection}
\subsection{Overview}
Anomaly Detection algorithms build models of normalcy for traffic patterns from historical motion data and utilize the models to identify vessels with anomalous characteristics. Ship anomaly can be classified into three types:
\begin{itemize}
\item	Position anomaly: a ship appears in a restricted/forbidden region or   in an unexpected position.
\item	Speed anomaly: The speed of a ship is significantly above or below regular speed in the same context.
\item	Time anomaly: The visiting time of a ship is unexpected.
\end{itemize}

The first two types are most commonly encountered and have been studied extensively. Different algorithms are capable of detecting different types of anomaly. In \cite{martineau2011maritime} anomaly detection methods are classified into three categories: statistical methods, neural networks and machine learning method. These categories significantly overlap as neural networks is a machine learning technique and many machine learning methods are statistical in nature. Here we categorize the anomaly detection algorithms into two classes, based on the models' learning characteristics:
\begin{itemize}
  \item Geographical (map-dependent) model based methods: build area-specific computational models which are trained on local traffic data and are superimposed on a geographical map of the locale to detect anomaly.
  \item Parametrical (map-independent) model based methods: build parametric models of normalcy independent of training region maps.
\end{itemize}


\subsection{Geographical (Map-Dependent) Models}
\subsubsection{Normalcy Box (NB)}
\begin{figure}[ht]
  \centering
  \includegraphics[width=5cm]{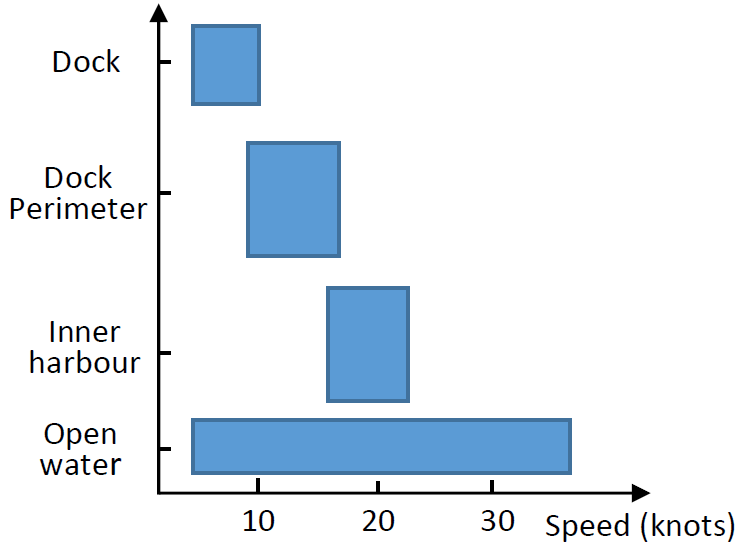}
  \caption{Learned normalcy box}\label{NormalcyBoxModels}
\end{figure}

The normalcy box method \cite{rhodes2005maritime} defines  a series of nesting regions over the port area according to the distance from a dock. Starting from the innermost part, the regions are: dock region, dock perimeter region, inner harbor region and open water region. The learning system is initially presented with a set of vessel's historical motion data containing location and corresponding speeds, labeled as normal/acceptable by a subject matter expert. The system then learns acceptable maximum and minimum speed for vessels in each region. The illustration in Fig. \ref{NormalcyBoxModels} demonstrates the simplest case with one normalcy box for each region, where regions are represented along the ordinate and vessel speed is along the abscissa. Each normalcy box is colored in blue and defines the spatial region and corresponding acceptable speed limits. For example, in the dock region, normal speeds are below 10 knots. A more complex model can be built to achieve better results by including more normalcy boxes in each region, as in the paper \cite{rhodes2005maritime}.

After the normalcy models (blue boxes in Fig \ref{NormalcyBoxModels}) are learned, anomaly detection is straightforward. Any observation that falls sufficiently far outside of the set of normal boxes is considered as unusual and an alert is raised.

In \cite{rhodes2009anomaly}, the normalcy box method is further improved by replacing the rectangles with hyper-ellipsoid, in order to have  a tighter boundary of the speed and space limit.

The advantages of the normalcy box approach lies in its online learning ability and its efficiency. Given new data, the normalcy models can be updated dynamically at any time in a very efficient way, either by adjusting existing blue boxes or by creating/deleting existing blue boxes. Anomaly detection can be achieved very quickly by just testing whether a location and speed pair falls within any blue box. The disadvantage of this method is that defining different regions is critical to detection results and nontrivial as it needs an accurate measure traffic distribution and prior expert knowledge.  Furthermore, there is no straightforward way to include pertinent static information like vessel type in the model. Such information may be differentiating factors in what constitutes normal behavior, e.g. different vessel types will have different normal speed.

\subsubsection{Fuzzy ARTMAP (FAM)}
The Fuzzy ARTMAP approach \cite{bomberger2006associative} first discretizes ships' course into four directions: north, south, east and west. On each direction, the speed is also discretized  into three states: slow, medium and fast. Then a latitude-longitude grid neural network is defined over the geographical area of interest, with the junctions constituting nodes in the neural network and the grid edges the synaptic connections.

The weight of a synaptic connection between node $i$ and $j$ is denoted as ${{w}_{ijk}}$, where $k$ is the index of speed state. When ship trajectory training data is presented, the weights in the neural network are updated according to the following equation:
\begin{equation}
 \Delta {{w}_{ijk}}=lr\cdot {{x}_{jk}}(t)\cdot \left( {{x}_{ik}}(t)-{{w}_{ijk}} \right)
\end{equation}
where $lr$ is the learning rate to be set manually, ${{x}_{ik}}(t)$ and ${{x}_{jk}}(t)$ are the start node (location) and target node (location) of the ship at time $t$, respectively. Learning is (presynaptically) gated by activation at the source node. If the source node is inactive, then weights at all synaptic connections originating from the node are unchanged. Otherwise, weights at synaptic connection linking the source node to any active target nodes are increased while weights at synaptic connections to any inactive target nodes are decreased.
The prediction of the vessel's future position is based on the weight of the connections originating from its current location. The larger the weight is, the more likely position the ship will be in the future. If a ship's route deviates from likely paths, it will be labelled as anomalous.

In \cite{rhodes2007probabilistic} the learning rule is further improved to make the learning rate self-adaptive with the following update rule:
\begin{equation}
  \Delta {{w}_{ijk}}=\frac{1}{{{N}_{jk}}}\cdot {{x}_{jk}}(t)\cdot \left( {{x}_{ik}}(t)-{{w}_{ijk}} \right)
\end{equation}
where ${{N}_{jk}}$ denotes the time instances node $j$ was activated.  According to the authors, the modification also lends learned network to a probabilistic interpretation. In \cite{zandipour2008probabilistic} the authors studied the effect of grid size upon prediction accuracy and proposed to use a multiscale network for a real-world problem, i.e. to define a dense network over busy and important regions and sparse network over other regions.

The strengths and weaknesses of FAM are as follows. FAM is an unsupervised learning method, i.e. it does not require learning data to be manually labelled as normal or abnormal. It is an efficient online learning method and can easily include new incoming data to improve its prediction accuracy. The prediction process is also computationally inexpensive.  However, it needs a large quantity of historical data to learn the weights of the networks and the prediction accuracy and path resolution are highly dependent on the size of the grids in the method. The grid size translates to the density of the network and needs to be tuned by experience for different regions. In addition, it is also non trivial to take into account pertinent static information, such vessel type and size, in FAM. This raises a major practical concern as such information is readily available and, from a practical point of view, useful for anomaly detection and path prediction.

\subsubsection{Holst Model}
Holst model\footnote{According to our survey, this type of cell dividing and modeling approach was first proposed by Holst A. in \cite{holst2003anomaly}, so in this paper we simply use Holst model to refer to this type of approach for convenience.}  \cite{holst2003anomaly, laxhammar2008anomaly} also defines grids over the port area. However, instead of learning connection weight, it attempts to model the velocity distribution in each grid cell by a Gaussian mixture model (GMM), which is described by the following equation.
\begin{equation}\label{CellDistribution}
  p(x|\lambda )=\sum\limits_{i=1}^{M}{{{w}_{i}}g(x|{{\mu }_{i}},{{\Sigma }_{i}})}
\end{equation}
where $x$ is a $D$ dimensional feature vector (i.e. location, speed, course etc.). ${{w}_{i}}$ is the mixture weight and $g(x|{{\mu }_{i}},{{\Sigma }_{i}})$  is the component Gaussian density function with mean vector ${{\mu }_{i}}$ and variance matrix ${{\Sigma }_{i}}$. The weights, mean vectors and variance matrices are the model parameters to be learned from training data, and are denoted as
\begin{equation}
  \lambda =\{{{w}_{i}},{{\mu }_{i}},{{\Sigma }_{i}}\}
\end{equation}
These parameters are estimated using Maximum Likelihood (ML) method, usually optimized by Expectation Maximization (EM) algorithm. By estimating the parameters of the mixture of Gaussians, the probability distribution of features is known and anomalous instances can be detected by thresholding.

Other learning methods may also be applied to estimate the probability distribution $p(x)$ in equation (\ref{CellDistribution}). In \cite{ristic2008statistical}, Ristic, B., \emph{et al} use kernel density estimation (KDE, also known as Parzen Window estimation) to learn the distribution of kinematic variables (position and velocity) in each cell. The results of these two methods are compared in \cite{laxhammar2009anomaly} and show that KDE can perform better than GMM.

The advantage of the Holst model is that it can adopt unsupervised learning methods to estimate a local density function (such as GMM, KDE), and therefore requires less human intervention. Because of the powerful fitting ability of GMM or KDE, the performance is usually very good. On the other hand, Holst model has high computational cost due to the estimation of probability in each cell; its prediction results varies depending on grid cell size/location and requires a large amount of normalcy training data for the probability estimation component to be reliable.

\subsubsection{Potential Field Method (PFM) }
Borrowing from the theory of potential fields in classical physics, the Potential Field Method (PFM) \cite{osekowska2013potential} as applied to anomaly detection instead of from AIS data models ship motion distribution as an electrical potential field and ship movement as a process of discharge. The general idea of PFM is that for the geographical traces of vessel movements, different amount of charges are assigned to all passed locations with respect to the number of ships and their speed. A collection of dynamically decaying charges distributed over an area generates a potential field, which is locally weaker or stronger depending on the density and strength of surrounding charges. Three main concepts of PFM are the total amount of a local charge, the decay of potential fields, and the distribution of a potential field around its charged source.

The total amount of a local charge is defined as a function of time and the accumulation of charges over a time period of $\tau $ is defined as follows:
\begin{equation}
 {{C}_{la{{t}_{k}},lo{{n}_{l}}}}(t)=\sum\limits_{t=0}^{\tau }{d(t)}{{c}_{la{{t}_{k}},lo{{n}_{l}}}}
\end{equation}
where ${{c}_{la{{t}_{k}},lo{{n}_{l}}}}$ is the component charge reflecting reported vessel properties: type, course, etc.; and $la{{t}_{k}},lo{{n}_{l}}$ are the geographical latitude and longitude coordinates at point $(k,l)$. $d(t)$ is a non-increasing decay function with limit at zero that describes the decrease of a local charge over time, for example $d(t)\propto 1/(t+\alpha)$ or $d(t)\propto \exp (-\alpha t)$ with $\alpha > 0$. Field distribution was implemented using the two-dimensional Gaussian smoothing equation \cite{jekeli1981alternative}. The local potential value is evaluated as
\begin{equation}
{{P}_{la{{t}_{k}},lo{{n}_{l}}}}(t)=\sum\limits_{i}{\sum\limits_{j}{\frac{1}{2\pi {{\sigma }^{2}}}{{e}^{-\frac{{{(la{{t}_{k}}-la{{t}_{i}})}^{2}}+{{(lo{{n}_{l}}-lo{{n}_{j}})}^{2}}}{2{{\sigma }^{2}}}}}}}
\end{equation}

A global potential field is instantiated by geographically distributed local charges. The intensity of the field varies depending on the geographic location and is determined by the strength of the surrounding local charges affected by their decay, and the distance to them. Areas associated with high potential  represent an emergent traffic pattern and describe a model of normal behaviour. Low or zero potential signify a lack of discernible normal traffic patterns in an area. An observed vessel behaviour that does not conform to the normal model described by the potential fields, is considered anomalous.

The performance of PFM at different grid size presented in \cite{osekowska2014grid} indicates that grid size has a significant impact on performance. An improved method implemented in \cite{woxberg2015maritime} allows for analysis of regions containing both low and high traffic density, which could not be done in the earlier work by Osekowska. This is achieved through the use of quadtrees to subdivide the grid used to create the potential field, allowing it to have an optimal grid size for different amounts of traffic in the same detection pass.

The most prominent feature of PFM is that it can detect both space and time anomaly using one single algorithm, because PFM takes both spatial and temporal pattern of trajectory (the charges) into account.  Another advantage is that the learned model (the potential field) is easy to be superposed on a geographical map, making it intuitive to users. The disadvantage of PFM is that it does not accommodate ship direction information in the formulation of normality. This is a significant drawback as sailing direction is particularly important where traffic lanes are separated. Additionally, ship type information is also disregarded in PFM.

\subsection{Parametrical (Map-Independent) Models}

\subsubsection{Trajectory Cluster Modelling (TCM)}
This method probably constitutes one of the first attempts at applying generic machine learning to anomaly detection in the maritime domain \cite{kraiman2002automated, riveiro2008supporting}. The points in a normal ship trajectory are represented by the vector $x_{k}^{t}=(lat,lon,speed,velocity);k=1...{{N}_{t}},t=1...T$ where $t$ is the index of different trajectory and $k$ is the index of the points in a trajectory. TCM combines the vectors $x_{k}^{t}$ and clusters them with a type of neural network called self-organizing map (SOM) \cite{kohonen1998self}. SOM produces a 2D plane with similar trajectory points gathered together and dissimilar points separated far away. A Gaussian Mixture Model (GMM) was then applied to model the characteristics of each cluster as a probability distribution.

The probability that a new sample is anomalous is obtained by applying Bayes rule on GMM probabilities \cite{roberts1998bayesian}. This allows the user to control the type 1 and type 2 errors by varying the anomaly detection threshold, a functionality that proved critical for operational applications. In addition to obvious anomalies, TCM also allows the detection of relatively small but persistently occuring anomalies by computing the probability of anomaly accumulated over a time window of operator-specified duration.

It is possible to replace both key components of TCM, i.e. SOM and GMM, by other clustering or probability modelling algorithm, such as kernel density estimation (KDE) modelling \cite{laxhammar2008anomaly}. The difference between Holst Model and TCM is that the former divided study area into small cells and models each cell, while the latter clusters all trajectories at once and model each class.

The advantage of TCM is that it provides a generic framework that can be easily extended to take into account relevant static information (size, shape, etc.) and kinematic information (location, speed, course, etc.). The performance is usually good because of the powerful modelling ability of GMM and nearly optimal Bayes estimation rule. The drawbacks of TCM are that it has high computational cost due to both the SOM and GMM components and it is not easy to update the models with a new incoming sample.

\subsubsection{Gaussian Process (GP)}\label{GProcessSection}
Gaussian Process (GP) has been demonstrated quite a powerful tool for both regression and classification problems \cite{rasmussen2006gaussian, kowalska2012maritime}. A Gaussian process defines a probabilistic distribution over functions, as opposed to Gaussian distribution(s), which define the probabilities of vectors. Suppose the aim is to learn an underlying mapping function $f:x\to y$ which is inferred from the posterior distribution
\begin{equation}
p(f)\tilde{\ }\mathcal{N}\left( \bar{f},V(f) \right)
\end{equation}
where $\mathcal{N}\left( \bar{f},V(f) \right)$ is a GP with mean function $y=\bar{f}(x)$ and variance function $V(f)$. $x$ is a vector containing ship position and $y$ is the predicted output, e.g. velocity. A GP is fully determined by its mean function and variance function. This is a natural generalization of the Gaussian distribution whose mean and covariance is a vector and matrix, respectively.

In \cite{kowalska2012maritime} GP regression models are constructed using training data $D={{\{x,y\}}_{1,...,n}}$ for the prediction of velocity ${{y}^{*}}$  at a new unseen position  ${{x}^{*}}$ . The prediction of ${{y}^{*}}$ is achieved through the construction of two separate GPs to predict ${{y}_{i}}$ (speed along horizontal direction) and ${{y}_{j}}$ (speed along vertical direction) velocity components. The mean function and variance function are given by
\begin{equation}\label{GPMean}
\bar{f}(x)={{k}^{*}}^{T}{{M}^{-1}}y
\end{equation}
\begin{equation}\label{GPVariance}
V(f)=K({{x}^{*}},{{x}^{*}})-{{k}^{*}}^{T}{{M}^{-1}}{{k}^{*}}
\end{equation}
where $M=(\tilde{K}+\sigma _{N}^{2}I),\ \ {{k}^{*}}={{\left[ K({{x}^{*}},{{x}_{1}}),...,K({{x}^{*}},{{x}_{n}}) \right]}^{T}}$ and $\tilde{K}$ is a kernel matrix generated with kernel function $K(x,y)=\exp \left( {-||x-y|{{|}^{2}}}/{2{{\sigma }^{2}}}\; \right)$. $\sigma _{N}^{2}$ is the variance of Gaussian noise produced during measuring process. In other words, equation (\ref{GPMean}) gives the expectation of the prediction value and equation (\ref{GPVariance}) gives the confidence about the predicted value.


Anomaly is detected based on a local anomaly score, which measures the deviation of the actual observation from the predictive distribution at each vessel position. The likelihood score is given by
\begin{equation}
score=\frac{1}{2}\log \left( 2\pi {{\sigma }^{*}}^{2} \right)+\frac{{{({{y}^{*}}-{{{\bar{y}}}^{*}})}^{2}}}{2{{\sigma }^{*}}^{2}}
\end{equation}

In \cite{will2011fast} the authors note that the mean function of GP can be written in a weighted sum form
\begin{equation}
\bar{f}={{k}^{*T}}{{M}^{-1}}y={{k}^{*T}}p=\sum\limits_{i=1}^{n}{{{w}_{i}}{{p}_{i}}}
\end{equation}
and this can be approximated by the Kd-tree which can significantly reduce GP computational burden.

The advantage of the GP method arises from its wide applicability and good track record in various fields. This can be very helpful when one considers to use it for anomaly detection application. The pleasant analytic properties of GP also mean that theoretical analyses can be readily performed. The main shortcoming of GP is its high computational cost and poor scalability, which remains a significant drawback for big data and/or real time applications, even with the numerous approximation algorithms for GP.

\subsubsection{Bayesian Network (BN)}
A Bayesian Network (BN) consists of a set of nodes (also called variables or attributes) $V=\left\{ {{v}_{1}},{{v}_{2}},...,{{v}_{|V|}} \right\}$, connected by directed edges $E=\left\{ {{e}_{1}},{{e}_{2}},...,{{e}_{|E|}} \right\}$  in a directed acyclic graph, $G$, where $|S|$ is the capacity of set $S$. Each node can take on a set of values (or states) which are typically discrete for computational purposes. The set of states for a variable is associated with a local probability distribution that is conditional only on the variable's parent nodes, i.e. the direction of the edge represents the conditional dependence relationship and the edge weight is the conditional probability. The network as a whole represents the joint distribution over its variables.

As illustrated in \cite{johansson2007detection}, to use Bayesian network, a set of variables and the corresponding states of each variable must be defined. For example one can define a variable representing ship type with the following states: $stat{{e}_{1}}\leftarrow $ cargo, $stat{{e}_{2}}\leftarrow $ passenger, $stat{{e}_{3}}\leftarrow $ tanker; or a variable representing ship speed with the states: $stat{{e}_{1}}\leftarrow $ (0-10 knots), $stat{{e}_{2}}\leftarrow $ (10-20 knots) and so on. These variables constitute the set of network nodes $V$. The directed edges in the network can be obtained from expert knowledge or by algorithmically identifying conditional dependence from training data using, for example, constraint-based algorithms and search-and-score algorithms \cite{nielsen2009bayesian}. There is also a practical toolbox to automatically construct the graph: CaMML \cite{korb2010bayesian}. After the edges are determined, the structure of the network is fixed. Values of the corresponding conditional probability distributions (i.e.  the edge weights) need to be estimated from training data. This can be easily accomplished once the structure of the network is fixed \cite{johansson2007detection}.

After the whole network is trained, anomaly detection is performed by computing the mean of $k$ consecutive joint probability of all variables over a time window
A vessel is flagged as anomalous if and only if the joint probability is below a threshold.

In \cite{mascaro2010learning} the authors build a more complex Bayesian network with a large number of variables, including kinematic information, weather condition, vessel details and vessel interactions. They also studied the network performance at two time scales: at individual time instances and throughout the track as a whole. Both results indicate that Bayesian network method is very promising for anomaly assessment and detection. More details about the algorithm are presented in a journal version clarify with of this paper \cite{mascaro2014anomaly}. The work presented in \cite{lane2010maritime} studied five anomalous ship behaviours: deviation from standard routes, unexpected AIS activity, unexpected port arrival, close approach, and zone entry. For each behaviour, a process is described for determining the probability that it is anomalous. Individual probabilities are combined using a Bayesian network to calculate the overall probability that a specific threat is present. Bayesian network is also widely used for anomaly detection in other fields, such as anomalous activity in computer vision \cite{loy2011detecting}, social network \cite{heard2010bayesian} and disease outbreak \cite{wong2003bayesian}.

The advantages of Bayesian network for anomaly detection are as follows: (1) A Bayesian network can easily incorporate expert knowledge into its structure. For example, a ship’s speed and type are usually highly related. When building a Bayesian network representation of vessel traffic, we can add a conditional dependence relation from the variable speed to ship type. (2) A fully defined Bayesian network is easily verified and validated by non-experts. This is very important in real-world applications where users are well acquainted with the application context but with little knowledge of the underlying algorithm. Such a user can very easily understand and verify a Bayesian network. (3) Bayesian networks are very powerful in modeling cause and effect and can easily include many influencing factors (such as weather, current, ship static and dynamical information) into one framework. The drawbacks of Bayesian network include the sensitivity of the performance to modeling assumptions like the variables selected and the definition of state, and the high computational cost. Extensive expert knowledge and a high level abstraction of the application context are usually needed to build high performing Bayesian networks.

\subsubsection{Other Methods}
Except for the above introduced methods, there are some other anomaly detection algorithms worth mentioning: SVM \cite{handayani2013anomaly}, Agent-System \cite{brax2009enhanced}, turning point detection \cite{vespe2012unsupervised}, TREAD system \cite{pallotta2013vessel}, POI/AP framework \cite{le2013unsupervised}, visualization influence \cite{riveiro2008improving, riveiro2009interactive,  riveiro2011role}.

\section{Route Estimation}
\subsection{Overview}
Route estimation refers to building up a model that captures the motion characteristics of a moving object and estimating the object's future position and trajectory from the model.

Comparing to other types of moving objects such as land vehicles and aircraft, ship motion is unique in the following ways: (1) A ship cannot abruptly stop, turn or reverse as a land vehicle does. It needs more time and space to transit from one motion state to another motion state. (2) For the practical purpose of navigation, a surface vessel movement occurs locally in a two dimension plane, whereas an aircraft or underwater vehicle motion occurs in three dimensional space. (3) In general, a ship typically has slow parabolic type maneuvers, while the fast-changing maneuvers are common in land and airborne vehicles. These unique characteristics differentiates motion modeling and prediction from other types of moving objects.

Methods of modeling and predicting vessel trajectory can be categorized into three classes according to their underlying implementation mechanism: physical model based methods, learning model based methods and hybrid methods. The first class models ship motion by a group of mathematical equations that precisely consider all possible influencing factors (mass, force, inertia, yaw rate, etc.) and calculate motion characteristics using physical laws. Because every factor is explicitly included in the modeling equations, once built the system can give  the ship exact trajectory in the future. The second class of methods model ship motion by one learning model that learns motion characteristics from historical motion such as is available from AIS data and thus implicitly integrates all possible influencing factors. Instead of considering all influencing factors explicitly, this type of methods treat the ship maneuvering system as a whole system and training the learning model using its historical data to mimic the system function, considering that its historical motion data are results produced under the effect of all influencing factors of the system. The third class of methods are hybrid methods which build a model that either explicitly considers part of influencing factors and is trained by historical motion data, or combines different learning methods together to form one model in order to have a better performance.

Physical models are very useful for building maritime simulation system for the purpose of training navigators or study ship kinematic characteristics. But they are rarely used independently for real-world ship trajectory prediction because, to perform well, such models need ideal environment and accurate state assumptions which are difficult to attain in reality. However, their ability to describe motion system can be useful when combined with machine learning methods.  There are many modelling methods proposed in ocean engineering area \cite{li2003survey}, but we will focus on  three most commonly used models: curvilinear model, lateral models and ship model. Learning model based methods for trajectory prediction are garnering more interest in recent years. Although many machine learning algorithms have been devised and successfully applied to various problems in the past 20 years, there is little work on applying machine learning to vessel trajectory prediction. In the third part of this section, we will discuss key methods including artificial neural networks, Gaussian Process, Kalman filter, support vector machine and minor principal component analysis.  Hybrid models aim to combine the strengths of its constituent models to improve prediction accuracy. We will introduce two typical combination types: Type I combines a physical model (e.g. curvilinear) with a learning method (e.g. Kalman prediction); Type II combines two learning methods together to improve performance.
\subsection{Physical Model Based Methods}
\subsubsection{Curvilinear Model}\label{CurvilinearSection}
\begin{figure}[ht]
  \centering
  \includegraphics[width=5cm]{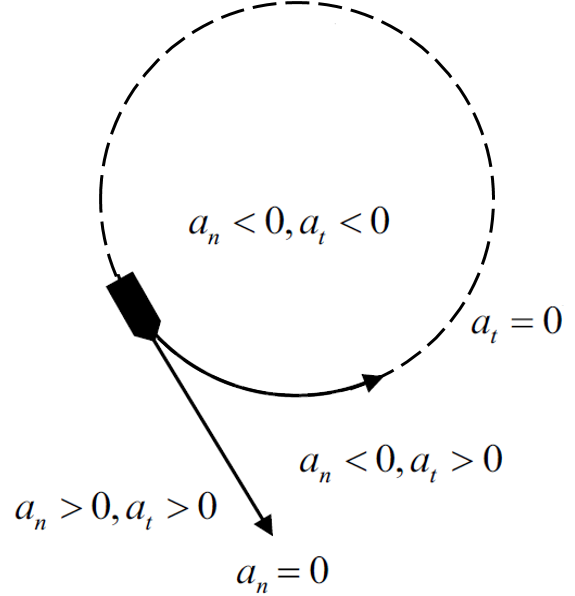}
  \caption{Curvilinear model}\label{CurvilinearModel}
\end{figure}

Curvilinear model \cite{best1997new} is a very general motion model that covers linear motion, circular motion and parabolic motion. As shown in Figure \ref{CurvilinearModel}, ${{a}_{n}}$ is normal acceleration and ${{a}_{t}}$ is tangential acceleration. Under the null normal acceleration, ${{a}_{n}}=0$, the model performs straight line motions; and when tangential acceleration, ${{a}_{t}}=0$, the model performs circular motions. Furthermore, the different acceleration conditions ${{a}_{t}}>0$ and ${{a}_{t}}<0$ produce various parabolic navigation trajectories.

The standard discretized system dynamic is described by the following equation \cite{li2003survey}
\begin{equation}\label{CurvilinearDynamics}
{{s}_{k+1}}={{F}_{cv}}{{s}_{k}}+{{G}_{k}}(s){{a}_{k}}+{{w}_{k}}
\end{equation}
where $s=\left( x,\dot{x},y,\dot{y} \right)$ and $a=\left( {{a}_{n}},{{a}_{t}} \right)$. $(x,y)$ are the target position in Cartesian coordinates. $\dot{x}$ means the derivative of $x$ with respect to time $t$. ${{w}_{k}}$ is white noise. ${{F}_{cv}}$ and ${{G}_{k}}(s)$ are two state transition matrices that describe the nonlinear motion property of the system.
Evaluation of matrix ${{G}_{k}}(s)$ involves nonlinear matrix integral which is hard to be computed precisely, but it can be approximated under some assumptions \cite{li2003survey}.

Kinematic models proposed for tracking a target moving in the horizontal plane can be  constructed with the following standard curvilinear-motion model from kinematics
\begin{equation}\label{CurvilinearKinematic}
\left\{ \begin{aligned}
  & \dot{x}(t)=v(t)\cos \phi (t) \\
 & \dot{y}(t)=v(t)\sin \phi (t) \\
 & \dot{v}(t)={{a}_{t}}(t) \\
 & \dot{\phi }(t)={{{a}_{n}}(t)}/{v(t)}\; \\
\end{aligned} \right.
\end{equation}
where $v,\phi $ are ground speed, and (velocity) heading angle, respectively, and ${{a}_{t}}$ and ${{a}_{n}}$ are the target tangential (along-track) and normal (cross-track) accelerations in the horizontal plane.

In \cite{perera2012maritime} curvilinear model is adopted in extended Kalman filtering to perform ship tracking and position prediction. In \cite{schubert2008comparison} the authors survey numerous (especially curvilinear) models and compares their performance in a tracking tasks which includes the fusion of GPS and odometry data with an Unscented Kalman Filter.

The advantage of curvilinear model is that it is a general model covers linear motion, circular motion and parabolic motion. These motion types are very common for ship and thus curvilinear model is particular useful for modelling ship motion. The disadvantage is that exactly solving equation (\ref{CurvilinearDynamics}) is very hard, especially matrix ${{G}_{k}}$. Various assumption has to be made in order to obtain an approximation of it.

\subsubsection{Lateral Model}
Any two dimensional motion can be decomposed in longitudinal direction and lateral direction. Lateral model \cite{huang2006vehicle, caveney2007numerical} (also known as bicycle model) focus on modeling lateral motion characteristics, since the longitudinal motion can be predicted with simple integration, with the longitudinal acceleration ${{a}_{x}}$ as the longitudinal input. Given steering angle $\delta $ as the lateral input, a general lateral model is described by
\begin{equation}\label{LateralModelEquation}
\left\{ \begin{aligned}
  & {{{\dot{v}}}_{y}}={{\beta }_{1}}\frac{{{v}_{y}}}{{{v}_{x}}}+\left( \frac{{{\beta }_{2}}}{{{v}_{x}}}-{{v}_{x}} \right){{\omega }_{z}}+{{\beta }_{3}}\delta ({{t}_{n}}) \\
 & {{{\dot{\omega }}}_{z}}={{\beta }_{4}}\frac{{{v}_{y}}}{{{v}_{x}}}+{{\beta }_{5}}\frac{{{\omega }_{z}}}{{{v}_{x}}}+{{\beta }_{6}}\delta ({{t}_{n}}) \\
\end{aligned} \right.
\end{equation}
where ${{v}_{x}}$ and ${{v}_{y}}$ are the longitudinal and lateral velocities. $\beta_1$ to $\beta_6$ are functions of ship intrinsic characteristics, such as mass, size, inertia, mass center, etc.
With ${{v}_{y}}$ and ${{\omega }_{z}}$ from the model, vehicle future trajectories can be predicted based on a simple geometric relationship \cite{huang2006design}, i.e.
\begin{equation}
\left\{ \begin{aligned}
  & \dot{x}={{v}_{x}}\cos \varphi  \\
 & \dot{y}={{v}_{x}}\sin \varphi  \\
 & \dot{\varphi }={{\omega }_{z}} \\
 & {{{\dot{v}}}_{x}}={{a}_{x}} \\
\end{aligned} \right.
\end{equation}
In \cite{pepy2006path} the lateral model is used in combination with Rapidly-exploring Random Tree (RRT) planner for vehicle path prediction and planning.

The lateral model is a simple and general model that covers constant steering, constant yaw rate and constant heading motion types. But the disadvantage is that parameter assumption and input assumption are critical and the measurement of the desired variables in equation (\ref{LateralModelEquation}) may not always be available in real applications.
\subsubsection{Ship Model}
Previous two physical models are general models applicable for land vehicles, aircraft and ships, even underwater vehicles. Ship model is specially designed for ship motion description and prediction. Ship model is a precise dynamic model which takes into account physical dimensions of the vessel and this is able to predict motion more accurately. There are many versions of ship dynamic models. The model proposed in \cite{pershitz1973ship,li2003survey} is particularly interesting as it involves fewer variables while being more generally applicable across vessel types. Its continuous time descriptive equations are
\begin{equation}
\left\{ \begin{aligned}
  & \dot{x}=v\sin (\phi -\beta ) \\
 & \dot{y}=v\cos (\phi -\beta ) \\
 & \dot{\phi }=K\Omega  \\
 & \dot{\Omega }=-\frac{v_{0}^{2}}{2p{{L}^{2}}}\left[ \frac{qL}{{{v}_{0}}}\Omega +{{s}_{32}}\delta  \right] \\
 & \dot{\beta }=-\frac{{{v}_{0}}}{2pL}\left[ q\beta +{{s}_{21}}\delta  \right] \\
 & v=K{{v}_{0}} \\
 & K={{\left( 1+\frac{1.9{{\Omega }^{2}}{{L}^{2}}}{v_{0}^{2}} \right)}^{-1}} \\
\end{aligned} \right.
\end{equation}
Here $(x,y),\phi  ,\Omega ,\beta ,\delta $ are ship position, heading, velocity vector turn rate, drift angle, and control ruder angle deviation, respectively; $v=v(\Omega )$ and ${{v}_{0}}=v(0)$ are ship speeds at turn rate $\Omega $ and $\Omega =0$ (i.e., at the onset of the turn), respectively; the hydrodynamic constants $p,q,{{s}_{21}},{{s}_{31}}$ depend on ship geometry and size, in particular, ship length $L$. This model has been used for ship tracking and position prediction in \cite{semerdjiev2000variable, semerdjiev1998adaptive}.

The advantage of this model is that given the parameters’ value, it can be used to predict ship trajectory in a more accurate way. While these parameters can be easily determined with high accuracy for own vessels, they are less readily available for other vessels encountered at sea.  Therefore, it might be more useful in simulation system rather than in a real world application.

\subsection{Learning Model Based Methods}
\subsubsection{Neural Network Method }
\begin{figure}[ht]
  \centering
  \includegraphics[width=5cm]{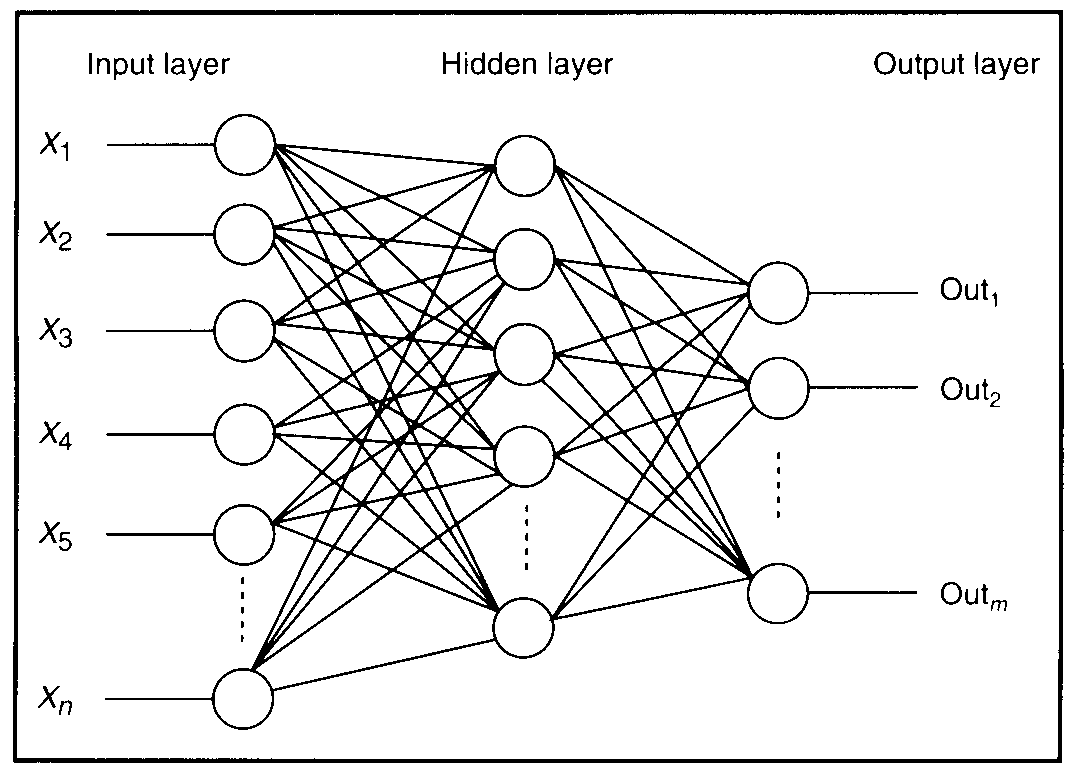}
  \caption{Basic structure of neural network model}\label{NeuralNetwork}
\end{figure}

Neural network \cite{haykin2004comprehensive} is one of the most popularly tools for regression due to its powerful ability of fitting complex functions. The basic structure of a multilayer feed forward network has three layers, the input layer, the hidden layer and the output layer. Neurons in different layers are connected by a weighted synapses and neurons in the same layer usually do not connect with each other in multilayer feed forward network, as shown in Figure \ref{NeuralNetwork}, in which the circles represents neurons and lines represents synapse. The input layer neurons receive input signals and transmit the signals to hidden layer neurons. The hidden layer perform computation and mapping its results according to its activation function to all output layer neurons. Each of output layer neurons finally sum up all its inputs to yield network output. Training a network means to adjust its synaptic weights so that the network output is close to a desired value.

Trajectory prediction using neural network usually consists of the following procedures:
\begin{enumerate}
  \item Define the mapping function from input to output ${{y}_{T+1}}=f(w,{{x}_{t}}|t=1...T)$ , where $w$ is the set of synaptic weights to be determined in training stage. ${{x}_{t}}$ is the feature vector containing ship static information (size, shape, weight, etc.) and kinematic information (location, speed, course, etc.). $T$ is the time length of the prediction period.
  \item Determine the network structure that used for the prediction task. This usually consists of choosing neuron activation function for hidden layer and output layer, determining neuron number in the hidden layer (input and output layers are automatically determined once the mapping function is fixed in previous step).
  \item Training the network using the training data set. Usually back propagation (BP) or Extreme Learning Machine (ELM) algorithm can be adopted. To make sure that the network has a good generalization ability, cross validation can be employed during training.
  \item Prediction of future trajectory. After the network is well trained, it can give future position while new feature vector $x$ is presented.
\end{enumerate}

There are several papers papers on trajectory prediction using neural networks. The differences between each other papers are the way they define the mapping function and the network structure. \cite{khan2005ship} defines the mapping function as
\begin{equation}
{{\phi }_{61,...,67}}=f({{x}_{t}}|t=1..60)
\end{equation}
and tries to predict ship course after 7 seconds, using the past 60 seconds location to train a network with 15 hidden neurons in the hidden layer. The authors also test the network performance for different input number and neuron number in hidden layer to verify the network performance. In \cite{xu2012novel} the authors define the mapping function as
\begin{equation}
{{[\Delta \phi ,\Delta \lambda ]}_{31}}=f({{c}_{t}},{{v}_{t}}|t=1,...,30)
\end{equation}
where $c,v,\Delta \phi ,\Delta \lambda $ are course, speed, difference of longitude and latitude, respectively. The authors tested networks performance with 4, 6, 8 hidden neurons and training epochs from 500 to 2000. Results show little difference, but the training algorithm has a significant influence (i.e. BP is much better than Mercator). In \cite{simsir2014decision} the authors construct a network to predict the following mapping
\begin{equation}
\left\{ \begin{aligned}
  & [{{x}_{g}}(t+3),{{y}_{g}}(t+3)]=f\left( {{x}_{g}}({{t}_{g}}),{{x}_{r}}({{t}_{r}}),{{y}_{g}}({{t}_{g}}),{{y}_{r}}({{t}_{r}}) \right) \\
 & {{t}_{g}}=0,-1,-2;{{t}_{r}}=1,2,3 \\
\end{aligned} \right.
\end{equation}
where ${{x}_{g}}$ and ${{x}_{r}}$ are the vessel actual coordinates and reference coordinates \cite{simsir2009prediction}. The network is trained with Levenberg–Marquardt learning algorithm. In \cite{zissis2015real} the authors use a network with 53 hidden neurons to predict the following mapping function
\begin{equation}
[la{{t}_{5}},lo{{n}_{5}}]=f(la{{t}_{t}},lo{{n}_{t}}|t=1..4)
\end{equation}
where $lat$ and $lon$ are the latitude and longitude of ship position. There are also other works to predict ship trajectory using neural network \cite{pietrzykowski2000prediction, jiehua2011nonlinear}.

The advantage of neural network method is that it is a general method and have been deeply studied in many areas, so its performance is usually stable and good. Also, there is no assumption/prior information (such as ship and weather), is needed, because its powerful fitting ability can theoretically learn any complex mapping function. The disadvantage is that the training process is usually very slow to convergence and there is no general rules about how to choose the activation function and hidden layer, so the network architecture need to be determined empirically according to the prediction task.

\subsubsection{Gaussian Process Method}
As described in Section \ref{GProcessSection}, Gaussian Process (GP) is a very powerful tool for prediction ship trajectories. The details of GP will not be described repeatedly. Please refer to Section \ref{GProcessSection} for its details.

In \cite{joseph2011bayesian,aoude2011mobile}, a motion pattern is defined as a mapping from location $({{x}_{t}},{{y}_{t}})$ to the distribution of trajectory derivatives $\left( \frac{\Delta {{x}_{t}}}{\Delta t},\frac{\Delta {{y}_{t}}}{\Delta t} \right)$. The authors use $p(f)=\mathcal{GP}(\bar{f},V(f))$ to learn the mapping from location to trajectory derivative (velocity)
\begin{equation}
\left( \frac{\Delta {{x}_{t}}}{\Delta t},\frac{\Delta {{y}_{t}}}{\Delta t} \right)=f({{x}_{t}},{{y}_{t}}|t=1...T)
\end{equation}


The mean and variance function of the GP are defined as
\begin{equation}
\left\{ \begin{aligned}
  & \bar{f}=0 \\
 & K(x,y,{x}',{y}')={{\sigma }^{2}}\exp \left( -\frac{{{(x-{x}')}^{2}}}{2w_{x}^{2}}-\frac{{{(y-{y}')}^{2}}}{2w_{y}^{2}} \right) \\
 & \quad \quad \quad \quad \quad \quad +\sigma _{n}^{2}\delta (x,y,{x}',{y}') \\
\end{aligned} \right.
\end{equation}

And the prediction at a new location $({{x}^{*}},{{y}^{*}})$ is given by
\begin{equation}
\bar{f}({{x}^{*}},{{y}^{*}})=K({{x}^{*}},{{y}^{*}},X,Y)K{{(X,Y,X,Y)}^{-1}}\frac{\Delta X}{\Delta t}
\end{equation}
where the expression $K(X,Y,X,Y)$ is shorthand for the covariance matrix $\Sigma $ with terms ${{\Sigma }_{ij}}=K({{x}_{i}},{{y}_{i}},{{x}_{j}},{{y}_{j}})$. In \cite{pallotta2014context} the authors use Ornstein-Uhlenbeck Processes to predict future position. Ornstein-Uhlenbeck Processes is a special type of GP with stationary property, which means its mean function and variance function do not change over time. This is a rather strict assumption in real application.

The advantages and disadvantages of GP are also discussed in Section \ref{GProcessSection}.

\subsubsection{(Extended) Kalman Filtering Method}
The Kalman filter uses a system's dynamics model (e.g., physical laws of motion), known control inputs to that system, and multiple sequential measurements (such as data from sensors) to form an estimate of the system's varying quantities (its state) that is better than the estimate obtained by using any one measurement alone. To use Kalman filter to estimate ship position, one need to define a system model and a measurement model:
\begin{equation}\label{KFSystemModel}
{{x}_{k}}=f({{x}_{k-1}},{{u}_{k-1}})+{{w}_{k-1}}
\end{equation}
\begin{equation}\label{KFMeasurementModel}
{{z}_{k}}=h({{x}_{k}})+{{v}_{k}}
\end{equation}
where ${{x}_{k}},\ \,{{u}_{k}}$ and ${{z}_{k}}$ are the system state, control input and measurement, respectively. For example, $x=(lat,lon,speed,course,\,\ acceleration)$ and $z=(lat,lon)$. ${{w}_{k-1}}$ and ${{v}_{k}}$ are the process and observation noises which are both assumed to be zero mean multivariate Gaussian noises.

The Kalman filter is a recursive estimator, which consists of two computational phases: prediction phase and update phase. The predict phase uses state from the previous time step to produce an estimate of the state at current time step. In the update phase,  current prediction is combined with current observation information to refine the state estimate. Details are omitted here but can be found in \cite{hamilton1994time, grewal2011kalman}.

 \cite{perera2012maritime, perera2010ocean} uses extended Kalman filter jointly with curvilinear motion model to perform ship trajectory estimation. This method can be treated as a hybrid model and will be elaborated in section \ref{TypeIHM}. \cite{ra2006real} uses Kalman filter to predict long-term ship rolling motion on  waved sea surface. In \cite{prevost2007extended} extended Kalman filter is utilized to predict the position of Unmanned Aerial Vehicle (UAV). Thus the measurement variable is $Z=(x,y,z)$, the spatial coordinates of the UAV. The state of the system is defined as $X=\left[ X_{r}^{T},X_{m}^{T},X_{y}^{T},X_{p}^{T} \right]$ where ${{X}_{r}}$ is the object setpoint (the speed, course and altitude), $X_{m}^{T}$ is the output of system motion model and${{X}_{p}}$ is the corresponding position coordinates vector. \cite{ammoun2009real} uses Kalman filter to predict land vehicle trajectory for collision avoidance.

The advantage of (extended) Kalman filter is that it is a well-studied classical method, and thus has many successful applications that can be utilized. The short time prediction accuracy is usually very good. The disadvantage is that the model initial state and model assumption are critical to achieve good prediction performance. For extended Kalman filter, the solution is generally not globally optimal because of the nonlinearity of the system model and measurement model.

\subsubsection{Minor Principal Component}
Minor Component Analysis (MCA)  has been demonstrated to be a good route estimation algorithm. It has similar mathematics as the Principal Component Analysis (PCA), except that MCA utilize the eigenvectors corresponding to the minor components (eigenvalues) \cite{peng2006new, bartelmaos2005fast}.  Consider the data matrix is $X$ and its autocorrelation matrix is
\begin{equation}
R=\frac{1}{n}X{{X}^{T}}
\end{equation}
The sorted eigenvalues $\{{{\lambda }_{1}},{{\lambda }_{2}},...,{{\lambda }_{n}}\}$ and corresponding eigenvectors $\{{{v}_{1}},{{v}_{2}},...,{{v}_{n}}\}$ of $R$ can be computed. Recall that PCA use the principal eigenvectors, i.e. the eigenvectors corresponding to the largest eigenvalues. MAC works in a similar way but use the minor eigenvectors. MAC prediction algorithm works as follows:
\begin{enumerate}
  \item Choose $k$ eigenvectors corresponding to the $k$ smallest eigenvalues;
  \item Arrange the eigenvectors column-wisely in a matrix $B$;
  \item Determine the prediction window parameter $t$ and partition matrix $B={{\left[ B_{t}^{T}\quad B_{n-t}^{T} \right]}^{T}}$ and the prediction sample data vector $x={{\left[ x_{t}^{T}\quad x_{n-t}^{T} \right]}^{T}}$;
   \item Solve equation ${{B}^{T}}x=0$ to get the prediction ${{x}_{n-t}}=-{{\left( {{B}_{n-t}}B_{n-t}^{T} \right)}^{-1}}{{B}_{n-t}}B_{t}^{T}{{x}_{t}}$.
\end{enumerate}
${{x}_{n-t}}$ is the prediction part and ${{x}_{t}}$ is a vector containing the past data. In \cite{zhao2004ship}, the method has been compared with neural network, autoregressive model and Wiener predictor and the results demonstrate it as a promising method for ship motion prediction. In \cite{BZ2005minor} MCA is used to prediction ship motion for landing forecast system.

The advantage of this method is its simplicity, i.e. it is easy to be understood and implemented. But it may have limited ability to model nonlinear motion, because component analysis (e.g. PCA) is a linear transformation and will yield decay results when applied to nonlinear distributed data.

\subsection{Hybrid Model Based Methods} \label{TypeIHM}
\subsubsection{Type I Hybrid Model }
This type of hybrid model is a combination of a physical model and a learning model. As an example, we introduce a combination of curvilinear model and extended Kalman filtering for route estimation. The curvilinear model introduced in Section \ref{CurvilinearSection} is able to describe common ship motion patterns. So it is natural to adopt this model as the motion model in extended Kalman filter in equation (\ref{KFSystemModel}). Recall that the kinematics of the curvilinear model is described by equation group (\ref{CurvilinearKinematic}). It can be reformulated as \cite{perera2012maritime, perera2010ocean}
\begin{equation}
\dot{x}(t)=f\left( x(t) \right)+{{w}_{x}}(t)
\end{equation}

Meanwhile, the measurement model in extended Kalman filter in (\ref{KFMeasurementModel}) is formulated as linear model since ship position at each time step is available.
\begin{equation}
z(k)=\left[ \begin{aligned}
  & {{z}_{x}}(k) \\
 & {{z}_{y}}(k) \\
\end{aligned} \right],\quad h\left( x(k) \right)=\left[ \begin{matrix}
   x(k) & 0 & 0 & 0 & 0 & 0  \\
   0 & 0 & y(k) & 0 & 0 & 0  \\
\end{matrix} \right]
\end{equation}
where ${{z}_{x}}(k)$ and ${{z}_{y}}(k)$ are measurements of $x$ and $y$ positions of the target vessel.

After defining the motion model (system model) and measurement model, the standard extended Kalman filtering steps can be used to perform tracking and prediction. It is possible to use other physical models to replace curvilinear model in order to have different learning ability.

The advantage of this hybrid model is that providing the modelling ability of curvilinear model, this method can predict several types of ship motion and, furthermore, can predict simultaneously the location, speed and acceleration with only the location data input. The disadvantage is that the initial system state assumption and the noise assumption have great influence on the algorithm performance. They have to be chosen carefully according to prior knowledge and experience.

\subsubsection{Type II Hybrid Model }
This type of hybrid model is a combination of different learning algorithms in order to have a better route estimation results. This combination type usually has two components: one for learning ship motion characteristics and the other for optimization of overall model performance. For example, the combination of a least square support vector machine (LS-SVM) and particle swarm optimization (PSO) \cite{zhou2010lssvm}; the combination of neural network (NN) and genetic optimization (GA) \cite{khan2005ship}; the combination of Kalman filter and neural network for marine target tracking \cite{stateczny2011multisensor, guo2009improved}. Here we take the LS-SVM and PSO combination as an example.

Least square support vector machine (LS-SVM) \cite{suykens1999least} is a variant of SVM, in which the inequality constraints are replaced by equality constraints, i.e.
\begin{equation}\label{LSSVMProblem}
\begin{aligned}
  & J(w,b)=\frac{1}{2}||w|{{|}^{2}}+C\sum\limits_{i=1}^{n}{e_{i}^{2}} \\
 & s.t.\quad {{w}^{T}}\phi ({{x}_{i}})+b={{y}_{i}}-{{e}_{i}},i=1..n \\
\end{aligned}
\end{equation}
These subtle changes make a significant difference. The original SVM is a quadratic optimization problem only for classification purpose, but problem (\ref{LSSVMProblem}) can be reformulated as a linear system and can be used for both classification and regression. For the detail technique of solving these problem, please refer to \cite{suykens1999least}. There are two critical parameters to be manually tuned in real application, the regularization parameter and the kernel width parameter for Gaussian kernel. Tuning these parameters is a nontrivial process and thus impractical for maritime navigational applications which require route estimation to be self-adjusted and real-time responded. To resolve this,  \cite{zhou2010lssvm} proposed to use Particle Swarm Optimization (PSO) to optimize these parameters in order to have a system tuning automatically with good generalization ability.


To optimize LS-SVM using PSO, the fitness function of PSO is the cross validation error of LS-SVM
\begin{equation}
f=RMS=\sqrt{\sum\limits_{i=1}^{L}{{{e}^{2}}}}=\sqrt{\sum\limits_{i=1}^{L}{{{({{d}_{m+i}}-d_{m+i}^{*})}^{2}}}}
\end{equation}
where $d_{m+i}^{*}$ is the prediction result of the $i$-th sample. Then PSO runs with two particles that correspond to the LS-SVM parameters $\gamma ,\sigma $ and when converges, the best values are the optimal parameters’ value. 

Comparing with uni-model method, this type of multi-model hybrid method can automatically select a group of optimal parameters and train a machine learning algorithm for good generalization ability, and thus need less human attendance. It is also demonstrated in some applications that multi-model methods can usually achieve better results. But the drawback is that the training process may be prolonged as the optimization component needs to run the learning component  many times in order to find a  (sub-) optimal solution.

\subsubsection{Other Methods}
There are also some other methods which can also be used for route estimation, including: ROT based \cite{last2014comprehensive}, stochastic linear system \cite{liu2011probabilistic}, quaternion-based rotationally invariant longest common subsequence (QRLCS) \cite{hermes2009long} and sequential Monte Carlo method \cite{lymperopoulos2010sequential}.

\section{Collision Prediction}
\subsection{Overview}
Collision risk assessment is a crucial step which directly determines whether the route planning procedure is invoked.  There are several concepts playing important roles for collision risk assessment.
\begin{itemize}
  \item Ship Domain (SD): The surrounding effective water area in which the navigator of a ship wants to keep clear of other ships or fixed objects.
  \item Own Ship (OS): A ship we directly control.
  \item Target Ship (TS): All other ship around own ship, sometime also called obstacles.
  \item Distance at the Closest Point of Approach (DCPA): the smallest distance between domains of own ship and target ship during the process of approaching each other.
  \item Time to the Closest Point of Approach (TCPA): The time costed to reach DCPA point at current maneuvering state.
\end{itemize}
Collision risk assessment is performed by either detecting possible violation of SD, or defining a risk index based on SD, DCPA and TCPA.

\subsection{Ship Domain}
Ship domain is the smallest safety region around a ship that allows navigator to take a timely action to avoid any potential collision. Any violation of the ship domain is interpreted as a threat to navigational safety and may cause a collision. So the definition of ship domain is not only very important for collision detection, but also is a collision risk assessment method. Generally, ship domain definition is affected by the following factors:
\begin{itemize}
  \item	Vessel shape and size: larger ship usually have larger domain in order to have more space for navigator to operate.
  \item	Ship speed and course: faster ship has larger domain in order to have more time for navigator to make decision.
  \item	Regional traffic density: In higher density traffic region, the ship domain shape may be irregular in order to take less space and meanwhile also keep ship safe.
  \item	Water current and weather condition: bad weather condition may cause navigator's judgement to be inaccurate and thus need large domain.
  \item	Navigator’s skills and experiences: operation and decision made by a less experienced and skillful navigator may increase collision risk and thus large ship domain is needed.
    \item	Other possible factors such as ship type, loading/draught etc.
\end{itemize}
So ship domain definition is an important and open problem. Existing ship domain can be casted into three types: simple domain, compound domain and learnt domain, ordered from simple to complex.
\subsubsection{Simple Ship Domain}
\begin{figure}[ht]
  \centering
  \subfigure[]{
  \includegraphics[width=2cm]{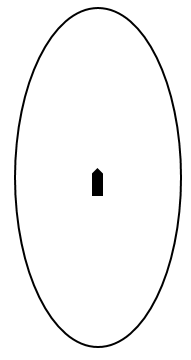}}
  \subfigure[]{
  \includegraphics[width=3cm]{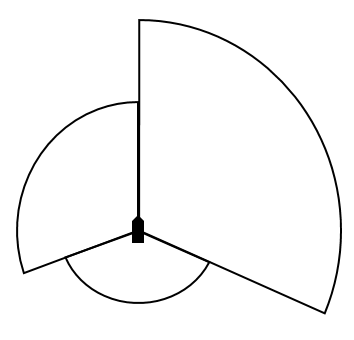}}
  \subfigure[]{
  \includegraphics[width=3cm]{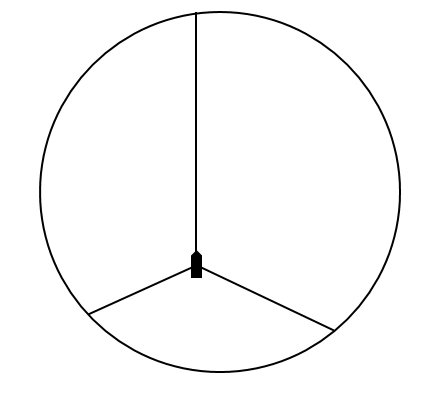}}
  \subfigure[]{
  \includegraphics[width=3cm]{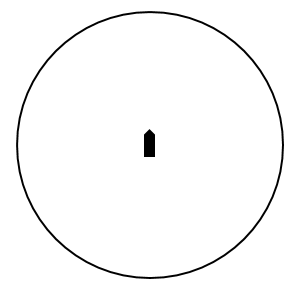}}
  \subfigure[]{
  \includegraphics[width=2cm]{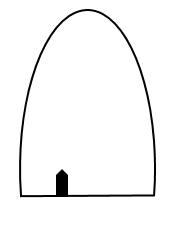}}
  \subfigure[]{
  \includegraphics[width=3cm]{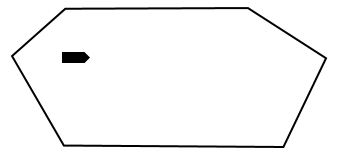}}
  \caption{Simple geometrical shape ship domains}\label{BasicShipDomain}
\end{figure}

This type of ship domain has a simple and regular geometrical shape. Figure \ref{BasicShipDomain} displays some simple ship domains that commonly used in various studies.
Fujii \cite{fujii1971traffic} first time proposed the concept of ship domain and gave a ellipse domain model in Figure \ref{BasicShipDomain}(a). Goodwin, E. M. \cite{goodwin1973statistical} studied the concept of ship domain and proposed a ship domain for open sea, as shown in Figure \ref{BasicShipDomain}(b). This domain emphasizes the front right area of the ship, because according to the International Regulations for Preventing Collisions at Sea 1972 (COLREGs) own ship is directly responsible for the risks between own ship and any target ship in this area. Contrarily, the front left area is much smaller, because own ship does not have direct responsibility. The rear part has the smallest domain region, since own ship usually moves forward and other normal target ship in this area has much less threat to the safety of own ship. The drawback of this domain model is that its boundary is not continuous and thus might be inconvenient for operation and computational simulation. In \cite{davis1982computer} a circle domain with off-centering own ship in Figure \ref{BasicShipDomain}(c) is used to approximate Goodwin’s domain model, in order to have a easily manipulated domain model. Hwang \emph{et al.} \cite{hwang2001design} use a circle shape domain in open sea in order to have a fast real time collision avoidance algorithm in Figure \ref{BasicShipDomain}(d). In restricted water, a compact ship domain should be employed in order to make use of the limited space. Coldwell \cite{coldwell1983marine} proposed to use a semi-ellipse in Figure  \ref{BasicShipDomain}(e) to model the ship domain for head-on and overtaking encounter situations in restricted waters. Mierzchalski \cite{michalewicz2000modeling} presented a more compact polygon domain in Figure  \ref{BasicShipDomain} (f) while study route planning. These domain models have simple shape and are defined based on experience. Once the domain is defined, the shape and size usually do not change anymore during all the sailing process, regardless the traffic situation or encounter situation. In \cite{wang2009unified} the authors developed an  analytic framework covering three main types of ship domain: circle \cite{goodwin1973statistical, davis1982computer} , ellipse \cite{fujii1971traffic, coldwell1983marine, pietrzykowski2012problem} and polygon \cite{michalewicz2000modeling, pietrzykowski2004ships, pietrzykowski2006ship}. Each type of ship domain has a uniform mathematical formula.

\subsubsection{Compound Ship Domain}
In previous section the ship domain models are a simple geometrical shape and their size and shape do not change once they are defined. This is usually not sufficient to keep ship safe in real application, in which the ship safety region is not always the same. For example, a navigator may want to have different different ship domain when a ship moves at different speed/course, or when sail from low traffic region to high traffic region. In these cases, compounded shape ship domains are developed to tackle these situations.

\begin{figure}[ht]
  \centering
  \includegraphics[width=3cm]{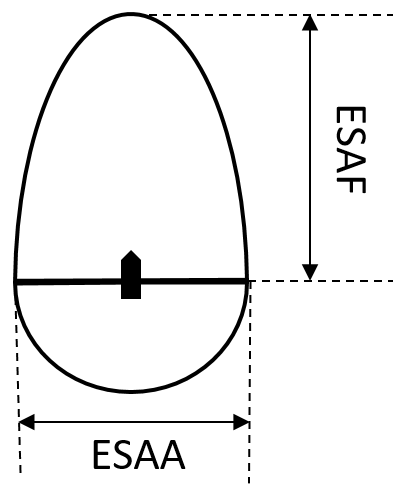}
  \caption{A compound ship domain with respect to ship speed}\label{CompoundShipDomain}
\end{figure}
In \cite{tam2010collision, tam2010path} the ship domain consists of a half ellipse at front and a circle at the rear. The major axis of the front ellipse and the radius of the rear circle change dynamically with respect to the ship’s speed, as shown in Figure \ref{CompoundShipDomain}. The front ellipse major axis is determined by
\begin{equation}
\text{ESAF=}\left\{ \begin{aligned}
  & {{V}_{TS}}S+D\text{,}\quad {{V}_{TS}}S+D\ge \text{MinSAD} \\
 & \text{MinSAD,}\quad \text{otherwise} \\
\end{aligned} \right.\
\end{equation}
where $V_{TS}$ is the speed of the target ship (TS) . MinSAD is the minimum distance that must be maintained between OS and TS for safety purposes. $D$ is a function of time step $\Delta t$. $S$ is a fixed value and generic scaling variable of the safety area that depends on the type of encounter (1.0min for ESAA and 1.5min for ESAF). The rear circle radius is determined by
\begin{equation}
\text{ESAA=}\left\{ \begin{aligned}
  & R+D\text{,}\quad \text{if}\ R\ \text{ }\!\!\times\!\!\text{ MinSAD}\ge \text{MinSAD} \\
 & \text{MinSAD,}\quad \quad \text{otherwise} \\
\end{aligned} \right.
\end{equation}
where $R$ is a function computing the safety area’s aft-section radius, and it is defined as follows:
\begin{equation}
R\text{=}\left\{ \begin{aligned}
  & {{V}_{TS}}S\text{,}\quad {{V}_{TS}}S\text{  SASLimit} \\
 & \text{2}\ \text{SASLimit - }{{V}_{TS}}S\text{,}\quad \text{otherwise} \\
\end{aligned} \right.
\end{equation}
where SASLimit is a predefined scalar property (typical 0.7 nmi)  that limits the maximum allowable safety area radius on the side and stern sections; it depends on the manoeuvrability of the TS.

In \cite{silveira2013use} a dynamical safety distance between ships are defined between two classes of ship (commercial; fishing vessels and pleasure craft) according to the encountered ship length, width and speed on separated traffic lanes
\begin{equation}
{{D}_{ij}}=L_{i}^{(1)}b+L_{j}^{(2)}a+B_{j}^{(2)}{{(1-{{a}^{2}})}^{1/2}}+B_{i}^{(1)}{{(1-{{b}^{2}})}^{1/2}}
\end{equation}
where $a=\left( {V_{i}^{(1)}}/{{{V}_{ij}}}\; \right)\sin \theta ;b=\left( {V_{j}^{(2)}}/{{{V}_{ij}}}\; \right)\sin \theta $,  ${{V}_{ij}}$ is the relative velocity between class $i$ and class $j$ vessels.
 $L_{i}^{(1)}$ is the length of class $i$ ships on waterway 1, $L_{j}^{(2)}$ is the length of class $j$ ships on waterway 2, $B_{i}^{(1)}$ is the breadth of class $i$  ships on waterway 1 and $B_{j}^{(2)}$ is the breadth of class $j$ ships on waterway 2. $V_{i}^{(1)}$ is the average velocity of class $i$ ships on waterway 1 and $V_{j}^{(2)}$  is the average velocity of class $j$  ships on waterway 2. The safety distance can be used to define ship domain and for collision risk assessment.

\subsubsection{Learnt Ship Domain}
Manually defined ship domain usually cannot take all the influencing factors into consideration. So some of the researchers consider to learn a ship domain from training data.

\begin{figure}[ht]
  \centering
  \subfigure[]{
  \includegraphics[width=4cm]{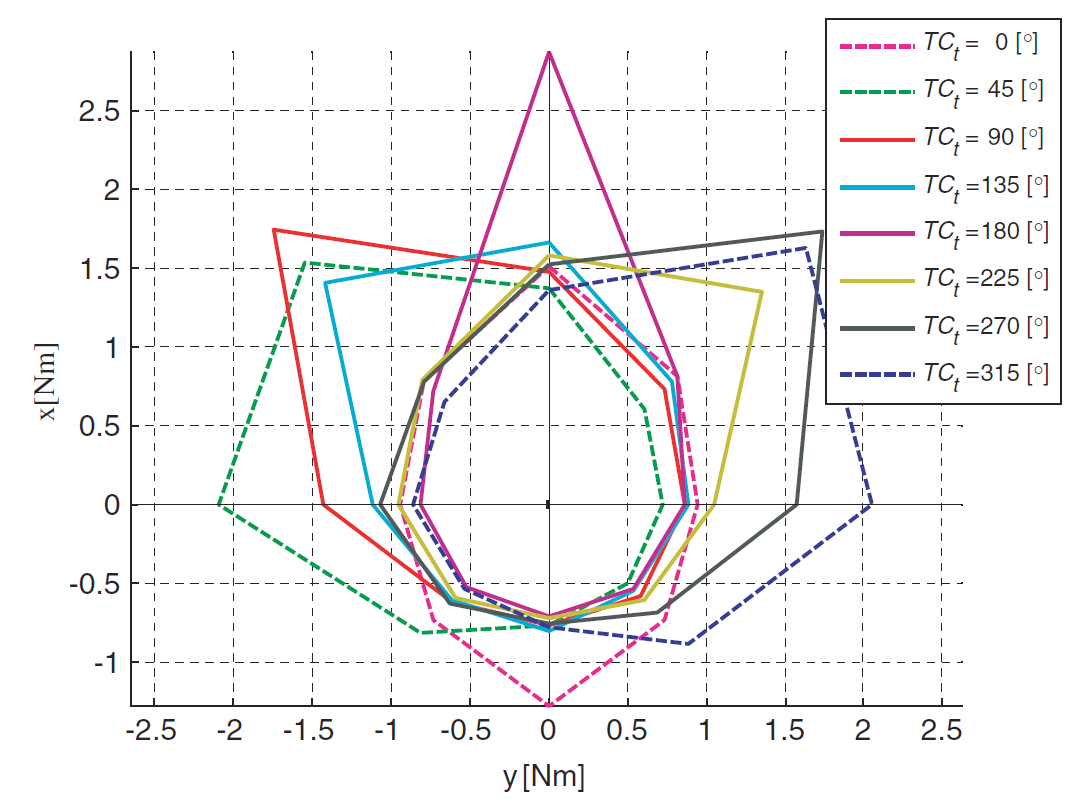}}
  \subfigure[]{
  \includegraphics[width=3cm]{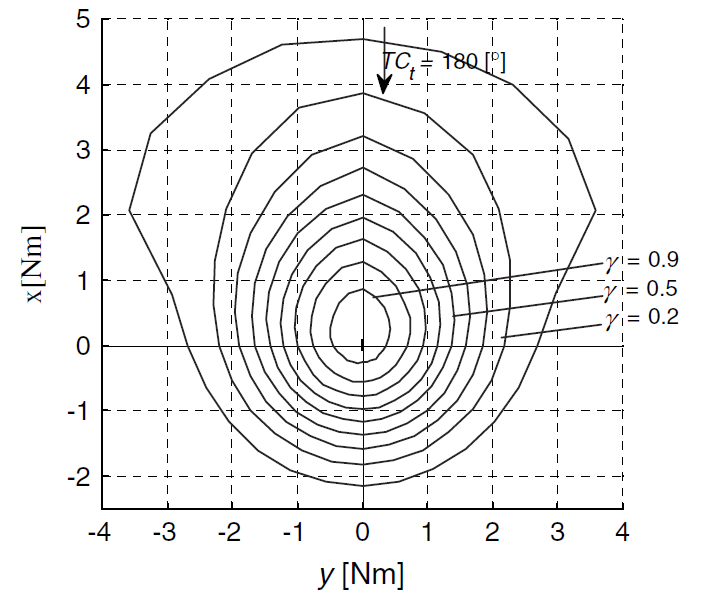}}
  \caption{Ship domain given by experts (a) and learning algorithm (b) (\cite{pietrzykowski2009ship}).}\label{LearnShipDomain}
\end{figure}

In \cite{pietrzykowski2009ship}, the authors conducted two questionnaires upon ship domain. The first questionnaire was on encountered ships with similar parameters and the second was on ships with different parameters. The participant were captains and watch officers with varying diverse sea experience. In questionnaires, the navigators were supposed to specify safety distances for various scenarios involving two ships: own and the target one.  Finally, the questionnaires results were used to train an artificial network (NN) to learn a ship domain of various size, encountering angle at different speed. Figure \ref{LearnShipDomain} shows the results of questionnaire (a) and the learnt domain  (b).

%

In \cite{lisowski2000neural} the authors constructed a neural network to learn a function
\begin{equation}
{{y}_{k}}=f({{u}_{k}})
\end{equation}
where ${{u}_{k}}=\left[ {{N}_{Bjk}},{{\mathbf{V}}_{k}},{{V}_{k}},\psi ,{{D}_{Abjk}} \right]$. The function value ${{y}_{k}}\in \left\{ 0.1,\ 0.3,\ 0.5,\ 0.7,\ 0.9 \right\}$ is the threat level where 0.1 represents safe and 0.9 represents collision. ${{N}_{Bjk}}$ a bearing to a target ship${{B}_{j}}$ , ${{V}_{j}}$ a speed of a target ship, $V$ speed of own-ship,  $j$ is the relative course of a target ship, ${{D}_{Abjk}}$ distance between own ship and target ship, and   $k$ is an index of a time moment.

In \cite{zhu2001domain} the authors also attempt to learn a ship domain model with a neural network, but their mapping function is different. The input of the network is
\begin{equation}
({{x}_{0}},{{x}_{1}},...,{{x}_{4}})={{\left( D/{{D}_{\max }},B/L,T/B,{{C}_{b}},\Phi /180 \right)}^{T}}
\end{equation}
where D is the visible distance, Dmax is valued to be 5 nm. B is the breadth and L is the length of own ship. T is the draft. ${{C}_{b}}$ is the block coefficient. $\Phi $ is bearing range. The target value of the network is $y=Dd/30L$ and it is the domain diameter.

In \cite{hansen2013empirical}, the authors study ship domain empirically according to a large amount of collected AIS data in a specific area when ships interacting with each other and give a estimated ship domain regarding to different influencing factors .

The advantage of learning a ship domain is that the learnt domain is flexible and self-adaptive, and can take different influencing factors into consideration. However, the disadvantage is that the model requires many training samples which need to be given by experienced navigators.

\subsection{Collision Risk Assessment}
\subsubsection{CPA Based Risk Index}
CPA means the closest point of approach, i.e. the ships’ locations that attain the smallest distance during the process of ships approaching each other. It is a crucial indicator of the ships’ collision risk and can be used to define various risk indices. In this section we introduce three risk indices that are based on CPA.

In \cite{hwang2001design, kearon1977computer} the collision risk index between own ship and the ${{i}^{th}}$ target ship is defined as a weighted sum of squares of DCPA and TCPA
\begin{equation}\label{CollisionIndexR1}
{{R}_{1}}(i)={{(aDCP{{A}_{i}})}^{2}}+{{(bTCP{{A}_{i}})}^{2}}
\end{equation}
where $a$ and $b$ are weight coefficients. $i$ is the number of target ship. This index is a rather simple one. When ${{R}_{1}}$ reaches a preset threshold value, the collision-avoidance action must be taken.

\cite{lisowski2001determining} further improve ${{R}_{1}}$ by including ship distance and normalizing all the term by safe factors
\begin{equation}\label{CollisionIndexR2}
{{R}_{2}}={{\left[ {{a}_{1}}{{\left( \frac{DCPA}{{{D}_{S}}} \right)}^{2}}+{{a}_{2}}{{\left( \frac{TCPA}{{{T}_{S}}} \right)}^{2}}+{{a}_{3}}{{\left( \frac{D}{{{D}_{S}}} \right)}^{2}} \right]}^{-\frac{1}{2}}}
\end{equation}
where $D$ is current distance between the own ship and the target ship, ${{D}_{S}}$is safe distance of approach (a radius of the circle-shaped domain), ${{T}_{S}}$ is the time necessary to plan and perform a collision avoidance manoeuvre, ${{a}_{1}},{{a}_{2}},{{a}_{3}}$ are weight coefficients, dependent on the state of visibility at sea, dynamic length and dynamic beam of the ship and a kind of water region. The advantage of equation (\ref{CollisionIndexR2}) over equation (\ref{CollisionIndexR1}) is that equation (\ref{CollisionIndexR2}) takes distance into consideration and thus it is dynamical and reflects current situation. Furthermore, the normalization factors are different for various types of ships, and thus equation (\ref{CollisionIndexR2}) is more ship-adaptive.

Based on ${{R}_{2}}$, \cite{szlapczynski2006unified} proposed a more generalized version by including ship speed, course, distance and ship domain into the risk index
\begin{equation}
{{R}_{3}}={{\left[ {{a}_{1}}{{f}_{{{\min }^{2}}}}+{{a}_{2}}{{\left( {{{T}_{{{f}_{\min }}}}}/{{{T}_{s}}}\; \right)}^{2}}+{{a}_{3}}f(t) \right]}^{-1/2}}
\end{equation}
where ${{f}_{\min }}$ is a generalized DCPA and ${{T}_{{{f}_{\min }}}}$ is a generalized TCPA
\begin{equation}
\left\{ \begin{aligned}
  & {{f}_{\min }}=\sqrt{-{{{B}^{2}}}/{4A}\;+C} \\
 & {{T}_{{{f}_{\min }}}}=-{B}/{4A}\; \\
\end{aligned} \right.
\end{equation}
$f(t)$ is the approach factor
\begin{equation}
f(t)=\frac{D(t)}{{{D}_{S}}}=\sqrt{A{{t}^{2}}+Bt+C}
\end{equation}
$A,B,C$ are functions of speed, course, distance and ship domain. The relation between them are ${{f}_{\min }}=\min f(t)$ and ${{T}_{{{f}_{\min }}}}=\underset{t}{\mathop{\arg }}\,\left( \frac{df(t)}{dt}=0 \right)$. The main advantage of ${{R}_{3}}$ is that it can be evaluated for any shape of ship domains of the encountered ships and give dynamical risk index over time.

In \cite{liu2006case} the risk degree of own vessel with target $t$ is defined as
\begin{equation}
R_4= \frac{1-{{EEC}^{2{{(DCP{{A}_{t}}-{{\lambda }_{DCPA}})}^{2}}}}}{2}+\frac{{{EEC}^{2TCPA_{t}^{2}}}}{2}
\end{equation}
if $DCP{{A}_{t}}<{{\lambda }_{DCPA}}$ and 0 otherwise.  $\lambda_{DCPA}$ the DCPA threshold and may have different values for different encountering situation. $EEC$ is a constant representing the ellipsoid eccentricity ratio.

\cite{chen2015research} proposes a collision risk evaluation function by including the DCPA, relative distance ($R$), TCPA, the azimuth from this vessel to target vessel $\triangle B$ and the
speed ratio K as the main factors of evaluation
\begin{equation}
\begin{aligned}
   &R_5=a_{DCPA}U_{DCPA}+a_{TCPA}U_{TCPA}  \\
  &  \qquad +a_RU_R+a_{\triangle B}U_{\triangle B}+a_KU_K
\end{aligned}
\end{equation}
where $U_{DCPA}$, $U_{TCPA}$,  $U_R, U_{\triangle B}$ and $U_K$ are fuzzy membership functions of the evaluation factors. $a_{DCPA}$, $a_{TCPA}$,  $a_R, a_{\triangle B}$ and $a_K$ are weighted coefficients.

Overall, from ${{R}_{1}}$ to ${{R}_{5}}$,  the collision risk indices become more and more general by including more and more influencing factors in the assessment. Meanwhile the complexity also becomes higher.

\subsubsection{Fuzzy Logic Method}
Fuzzy logic is an extension to traditional Boolean logic, in which value of a variable is either true or false (1 or 0, corresponding). Fuzzy logic uses multiple input value, i.e. a variable may have many values varying from false to true, usually determined by a fuzzy membership function.

In \cite{hwang2001design}, Hwang et al proposed a fuzzy collision-avoidance system which consists of 5 fuzzy function modules: object detection module, static/moving object avoiding module, trajectory-tracking module and speed control module. The fuzzy rules of each module are determined according human intuitions and a $H_{\infty}$ autopilot system is developed to obtain optimal control output from the exogenous input (such as speed, draft, the depth of water, the encountering situations and the surrounding currents, winds and waves, etc).

In \cite{lee2004fuzzy}, Virtual Force Field (VFF) is combined with fuzzy roles to design collision avoidance system. Own ship motion is assumed to be dominated by the total force of two component forces: the attractive force ${{\vec{F}}_{a}}$ which pulls ship to target track and the repulsive force ${{\vec{F}}_{r}}$ which directs ship away from obstacles. The contribution of each force is determined by fuzzy rules of linguistic variables between own ship and target ship: relative distance, relative course and relative speed.

\begin{figure}[ht]
  \centering
  \includegraphics[width=4cm]{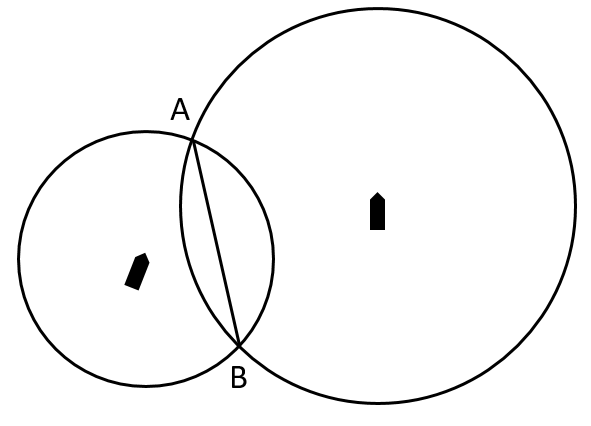}
  \caption{Guide rings and radical axis (chord $AB$)}\label{GuideRing}
\end{figure}

In \cite{kao2007fuzzy}, a collision risk assessment system is developed based on fuzzy logic method. The authors first defined a fuzzy ship domain (guarding ring) whose radius is determined according to three fuzzy variables: ship length L, speed V and sea condition S. The combination of the three linguistic variables has a total of 27 fuzzy rules that are used to determine the size of the guarding ring.
%
The radical axis is a line passing through the intersection points of two guarding rings, as shown in Figure \ref{GuideRing}. The length variation of the radical axis is considered as the collision alert index.  While two guarding rings are overlapping, the alert index function begins operation and a danger index is calculated as
\begin{equation}
{{\mu }_{danger}}(\Delta t)=\left\{ \begin{aligned}
  & 1,\quad \quad \quad \quad \quad \quad \Delta t=0 \\
 & 1-2{{\left( \frac{\Delta t}{\tau } \right)}^{2}},\ 0\le \Delta t\le \frac{\tau }{2} \\
 & 2{{\left( \frac{\Delta t-\tau }{\tau } \right)}^{2}},\ \frac{\tau }{2}\le \Delta t\le \tau  \\
 & 0,\quad \quad \quad \quad \quad \quad \Delta t\ge \tau  \\
\end{aligned} \right.
\end{equation}
where $\tau $ is obtained by a fuzzy model of the $S$ function and $\Delta t$ is the time discrepancy of two observations of the potential collision point. According to the value of the danger index, the two ships maintain a safe condition as long as the discrepancy of collision point $\Delta t$ for the two target ships is larger than $\tau $. Otherwise, the ships face a risk of collision.

In \cite{perera2011fuzzy}, own ship's domain is divided into 8 sectors with respect to different encountering situations (overtaking, cross and head on) according to COLREGs rules. In each sector,  collision risk is studied with respect to target ships course and position. To do so, the relative sailing information (position, speed and course) of a target ship is estimated and fuzzified with respect to collision distance, collision region, relative speed ratio and relative collision angle. Then an if-then fuzzy rule table is used to determine collision warning risk warning and fuzzy decisions. If potential collision exists, the fuzzy decision will be further defuzzified to obtain own ship speed and course adjustment to avoid the potential collision.
\section{Path Planning}
\subsection{Overview}
After collision risk assessment, if a potential collision is detected, own ship should take some action to avoid the potential collision. Path planning is to find the new safer route for own ship that has minimum cost (w.r.t time, distance, course changes etc) to eliminate collision risk. Traditional path planning is conducted manually by experienced navigators. The processes is laborious and time-consuming  and the route might be suboptimal. In contrast, an intelligent path planning system can take all possible influencing factors (such as traffic density, weather condition, encountering situation, etc.) into consideration to plan a optimal collision-avoidance path and meanwhile significantly reduce reaction time and human workload \cite{cummings2010supporting}. In this section we introduce four types of own ship path planning algorithms reported in the literature: shortest graph path method, evolutionary algorithm method, evolutionary set method and artificial potential field method.
\subsection{Shortest Graph Path Method}
Shortest graph path method aims to find the shortest path between two nodes on a graph, also known as maze routing \cite{chang2003method}. The original goal of maze routing problems \cite{lee1961algorithm} is to find a shortest path between a given pair of cells on a rectangular grid of cells without crossing any obstacles.

In \cite{chang2003method} the authors proposed a method to plan route for own ship to avoid moving and static obstacles based on a 4-geometry maze routing algorithm, which is an improved algorithm to the Lee’s algorithm \cite{lee1961algorithm, fagerholt2000shortest} to include higher geometry grid connection. The maze routing algorithm first assigns four variables to each cell: SL (sea or land indicator), AT (arrival time from source), SD (ship domain indicator) and Vis (visited indicator). Then the algorithm visits unvisited cells continuously according to breadth first rule until reaching destination cell to find out the shortest route between current position and destination  position. The advantage of their maze routing algorithm, comparing with popular Dijkstra’s algorithm, is that it does not need to construct an adjacency matrix and has linear complexity. Based on this maze routing algorithm, the authors proposed a shortest path collision avoidance scheme as follows:
\begin{enumerate}
  \item obtain the shortest path for each ship using their maze routing algorithm;
  \item Simulate all of the routes with ship domains;
  \item 	If any cell is visited twice, recompute the path for give-way ship and return to step 2);
  \item Give optimal route for each ship
\end{enumerate}

 One drawback with this path planning scheme is that when there are many obstacles (such as port area where there may be many ships), the shortest path algorithm tend to find a path that contains too many turning points. Ship turning is not only time consuming, but also much likely to cause danger, especially for sharp turning. Therefore an optimal path should contain as few turning points as possible. In \cite{szlapczynski2006new} the authors further improve the above maze routing algorithm by including a turning penalty in the AT variable. Specifically, the AT is replaced by GAT, which is defined as
\begin{equation}\label{GAT}
\begin{aligned}
  & GA{{T}_{new,j,gn}}=\min \left\{ GA{{T}_{i,1}}+{{d}_{i,j}}+{{t}_{gn,1}},...,+ \right. \\
 & \quad \quad \quad \quad \quad \quad \quad \quad \left. GA{{T}_{i,8}}+{{d}_{i,j}}+{{t}_{gn,8}} \right\} \\
\end{aligned}
\end{equation}
where: $d$ is the distance between two cells and $t$ is the corresponding sailing time cost. $i$ and $j$ are indices of the neighbouring cells. $gn$ is the current gate of the ${{c}_{i}}$ cell and numbers from 1 to 8 denote all gates of the ${{c}_{i}}$ cell.  $t$ is equal to zero for two gates of the same direction and has appropriate parameter values ${{d}_{1}},{{d}_{2}}$ or ${{d}_{3}}$ for two gates whose direction difference is 45, 90 or 135 degrees respectively. Therefore, equation (\ref{GAT}) tends to put large penalty on direction change.

In \cite{blaich2012fast, hornauer2015trajectory,cummings2010supporting}, another shortest path algorithm A* \cite{hart1968formal} is used to plan a path to avoid collision. Their method additionally regards the physical constrains of the vessel and the COLREGs. The difference between A* and Lee’s algorithm (or Dijkstra algorithm) is that the former is a depth-first searching algorithm which can perform faster on large graph but may results sub-optimal path; but the latter is a breadth-first searching algorithm which suitable for small and medium size graph and can give global optimal path.

The advantage of shortest path method is that it is simple and easy to be implemented and can guarantee to produce an optimal path if exists. The disadvantage is that it need to construct rectangular grids over study area.  Another disadvantage is that it requires the exact position/size of each obstacle and the destination information of each ship in order to run maze routing algorithm, but these information is usually not easy to be obtained.

\subsection{Evolutionary Algorithm Method}
Evolutionary algorithm (EA) is a powerful optimization tool that can simultaneously optimize multi-objective function subject to multi-constraints. Theoretically it can find a global optimal solution for any optimization problem, given sufficient population size and generation number. The key steps for route planning using EA are (take Genetic Algorithm, GA, as a example) :
\begin{enumerate}
 \item 	Encoding individual gene and chromosome: this step transforms a ship path to an individual chromosome in the GA. A common encoding method is to set each coordinate pair $(x,y)$ as a gene and then an individual chromosome is a series of genes that represent a series of these coordinate pairs $({{x}_{i}},{{y}_{i}}),i=0..n$  being connected to form a path.
 \item 	Define objective function and constraints: this step defines how an individual chromosome can survive and its superiority in the generation. Usually a chromosome (i.e. a ship path) with smaller distance and less time cost is likely to have larger chance to survive and produces descendants.
 \item Set parameters and run GA optimization: this step is to find out an optimal individual that best fits the objective function and meets all the constraint requirements. This step is a common step for all GA based route planning and it consists of several sub-steps:
     \begin{itemize}
 \item Initial population generation
 \item Fitness and constraints evaluation
 \item Survivals and elites selection
 \item Gene operations: Crossover, mutation, variation
 \item Reproduction, go to step 2 if not terminate
 \end{itemize}
\end{enumerate}

The differences between different EA based path planning methods are mainly in the first two steps, while the third step is a common step which has very little difference.
In \cite{ito1999collision} a chromosome is encoded as a line series and the objective function is defined as a sum of four terms
\begin{equation}
J=\alpha a+\beta b+\gamma c+\delta d
\end{equation}
where $\alpha ,\beta ,\gamma ,\delta $ are weight coefficients. $a, b, c, d$ are the level of danger, length of path, straightness of the route and energy loss, respectively.
This route planning method is further improved in \cite{zeng2000collision} by including a traffic navigation rule term in the cost function, namely, own ship must pass from the right of other moving ships. In \cite{zeng2003evolution} a noise term modelling the influence of tide, wind, and wave is introduced into the gene structure to achieve a better prediction while the ship sails in bad weather condition.

In \cite{smierzchalski1999evolutionary} a general planning objective is proposed for evolutionary algorithms with the cost function of a path $S$ being formulated as
\begin{equation}
J(S)=safe\_Cost(S)+Econ\_Cost(S)
\end{equation}
where the safety cost is
\begin{equation}
safe\_Cost(S)={{w}_{c}}\max \{{{c}_{i}}\}_{i=1}^{n-1}
\end{equation}
${{c}_{i}}$ is the length difference between the distance to the constraint-closest turning point ${{s}_{i}}$ and the safe distance. And economy cost
\begin{equation}
Econ\_Cost(S)={{w}_{d}}*dist(S)+{{w}_{s}}*smooth(S)+{{w}_{t}}*time(S)
\end{equation}
where ${{w}_{d}},{{w}_{s}},{{w}_{t}}$ are weight coefficients and $dist(S)$ is the total distance, $smooth(S)$ is the maximum turning angle and $time(S)$ is the sailing time of the path. This objective is implemented in \cite{michalewicz2000modeling} and further improved by including static (location) and dynamic information (velocity and speed limitation at turning point) into gene.

Another GA based path planning algorithm finds an alternative collision-avoidance path with minimum restoration cost to current path \cite{tsou2010decision}. In their algorithm, a chromosome contains four genes:
\begin{itemize}
  \item $Q$: The required time to the turning point (or the time from TCPA)
  \item $C{{'}_{o}}$: The required collision avoidance angle for passing the target ship at safe distance
  \item $Ta$: The time between the turning to collision avoidance and the turning to navigational restoration
  \item $Cb$: The limited angle upon turning of navigational restoration
\end{itemize}
The cost function is defined as
\begin{equation}
J=\min \{D{{s}_{i}}+D{{r}_{i}}\}_{i=1}^{n}
\end{equation}
where $D{{s}_{i}}$ is the distance after collision avoidance, $D{{r}_{i}}$ is the distance of navigational restoration. The constraints are turning angle ([30, 90] for avoidance and [-60, -30] for restoration) and restoration time (less than 60 min). Because the chromosome has fewer genes, their method can achieve a high execution speed.

There are also some other EA methods, such as \cite{cheng2007trajectory} uses GA to optimize the parameters in a mathematical motion model and after optimization uses the model to planning ship path. \cite{tam2010path} includes COLREGs into the optimization process and use GA to optimize the turning cost. \cite{ying2007ship} uses Bayesian rules to compute the posterior probability of each trajectory during GA optimization process. The colony algorithm in \cite{tsou2010study, lazarowska2014safe}.

The advantage of EA is that it can easily include various static and moving obstacles into constraints and can solve multi-target planning problem. However, EA usually need high computational cost and converges slowly. In real time application, this might be a big problem.

\subsection{Evolutionary Set Method}
In general, Evolutionary set (ES)  can also be treated an variant of evolutionary algorithm, but the difference from traditional evolutionary algorithms (such as GA or ant colony) is that ES does not have the encoding and decoding process. In other words, all evolution operations are directly defined on ship sailing path.

In \cite{szlapczynski2011evolutionary}, each individual (a population member) is a set of trajectories which corresponds to ships involved in an encounter. A trajectory is a sequence of nodes which contain geographical coordinates $x$ and $y$, the speed between the current and the next node. After initially generated some individuals, the authors define various evolutionary operators that directly applied to trajectories.
\begin{itemize}
  \item Reproduction: parental trajectories are crossed to produce new trajectories (offspring).
  \item Mutation:  randomly node insert, node joining, node shift and node delete on single trajectory.
  \item Special operators: specialised operations for improving trajectories and convergence rate.
  \item Validations and fixing operators: evaluate surviving trajectories and make adjustment for better applying to real application.
\end{itemize}	

The overall individual fitness is defined as a sum of its trajectories fitness
\begin{equation}
fitness=\sum\limits_{i=1}^{n}{\left[ {{tf}_{i}} \right]}
\end{equation}
where ${{tf}_{i}}$ is the fitness of each trajectory, defined over both moving ship and  static obstacles.

\cite{szlapczynski2012evolutionary} further improves the ES to an Evolutionary Sets of Safe Ship Trajectories (ESoSST) system, in which the optimization criterion and existing penalties were modified to include additional COLREGs-violation penalties. Most of the evolutionary mechanisms were extended or replaced with more advanced ones to improve the ESoSST method performance. \cite{szlapczynski2013evolutionary} introduces an extended ESoSST methodology, with a focus on  rule 10 in COLREGs and fully support Traffic Separation Scheme (TSS), including detecting and penalizing TSS violations, as well as the pre-processing phase (generating the initial population, which includes predefined TSS-compliant tracks). \cite{szlapczynski2015evolutionary} extends the previous EA to deal with restricted visibility, rule 19 in COLREGs.

Because the operations are directly applied to trajectories and no encoding/decoding process is needed, ES has much less computational cost and is reported to be much faster than traditional evolutionary algorithms. Therefore it can be used to search optimal trajectories for all encountered ships at the same time, not only just for own ship as the GA does.

\subsection{Artificial Potential Field}
The artificial potential field (APF) method was first introduced by Khatib \cite{khatib1986real} for robot path planning in the 1980s. The basic concept is to fill a robot’s workspace with an artificial potential field, in which the robot is attracted to its goal position and repulsed away from the obstacles.

In \cite{xue2009automatic,xue2011automatic}, the authors use potential field to plan ship path for collision avoidance in a simulation system. The ship sails towards a target point D along the gradient descent direction of the total potential field
\begin{equation}
U(p)={{U}_{att}}(p)+{{U}_{rep}}(p)
\end{equation}
where $p$ denotes a point on the water surface. ${{U}_{att}}$ is the potential energy owing to attraction towards destination point and is defined as
\begin{equation}
{{U}_{att}}(p)=\alpha {{\left| {{p}_{d}}-p(t) \right|}^{m}}
\end{equation}
 ${{U}_{rep}}$ is the potential energy owing to repulsion of the obstacle and is defined as
 \begin{equation}
{{U}_{rep}}(p)=\left\{ \begin{aligned}
  & \frac{1}{2}\eta {{\left| \frac{1}{{{p}_{s}}}-\frac{1}{{{p}_{o}}} \right|}^{2}}{{\left| p(t)-{{p}_{d}} \right|}^{n}},\text{if}\quad {{p}_{s}}\le {{p}_{o}} \\
 & 0,\quad \quad \quad \quad \quad \quad \quad \quad \quad \ \ \text{if}\quad {{p}_{s}}>{{p}_{o}} \\
\end{aligned} \right.
 \end{equation}
where ${{p}_{d}}$ and $p(t)$ denote the destination position and the position of ship at time $t$, respectively. The total virtual force exerted on the ship in this potential field can be calculated as
\begin{equation}
F(p)=\nabla U(p)=\nabla U{{(p)}_{att}}+\nabla U{{(p)}_{rep}}
\end{equation}
The ship motion is controlled under the force $F$ to sail toward target and avoid obstacles.

\cite{zhang2007autonomous} proposed a improved APF algorithm by including ship velocity and maneuvering behavior into the corresponding virtual force term. In \cite{shi2007harmonic}, to overcome the local minima shortcoming of APF, the authors proposed to construct a harmonic potential field for autonomous navigation
of ships in diverse environments. With proper boundary assumptions, the harmonic field satisfies Laplacian equation in potential domain and can eliminate local minima. Potential field method has been widely used for robot motion planning \cite{ge2002dynamic, jaradat2012autonomous}.

The advantage of potential field method is that it can provide dynamical route at every moment and thus the real-time capability is favourable. However, the calculation need to be performed at each time step and thus the optimal sailing course needs to adjust too frequently. For ship this may cause problem, because frequently changing course is time consuming and dangerous and should be avoid in sailing. Another problem is that it is not easy to produce the exact theoretical amount of driven force for a ship. Furthermore, modelling repulsive forces is a difficulty in reality and the resultant total force is sensitive to repulsive forces.

\subsection{Other Methods}
Besides the introduced methods, there are some other path planning algorithms/methods that are worth mentioning, including game theory \cite{lisowski2007dynamic, lisowski2012game}, fuzzy system \cite{hwang2001design, lee2004fuzzy}, maximin method \cite{miele2004new,miele2005maximin,miele2006optimal}; \cite{benjamin2004colregs} studies the COLREGs rules in detail and proposes a method of multi-objective optimization, interval programming (IvP) to determine ship behavior.
\section{Conclusions and Discussions}
 This paper surveys AIS data sources and four relevant aspects of navigation in which such data is or could be exploited for safety of seafaring, namely traffic anomaly detection, route estimation, collision prediction and path planning. From these surveys we can conclude as follows:
\begin{itemize}
\item For anomaly detection, generally speaking, geographical models are intuitive and easily visualized, but they usually use only part of available information and not easy to include domain knowledge or expert knowledge. In contrast, parametrical models are easy to include all useful information and expert knowledge, but they lack of intuitiveness and direct visualization.  Overall, all the methods are capable of detecting one or two types of anomalies, but hardly can detect all types. In addition, situation awareness and anomaly detection usually work in a parallel and iterative way, i.e. one improves the other.
\item Route estimation is a challenging task. In this paper we surveyed a broad range of methods developed for ship route estimation and categorize them into three classes. Generally speaking, physical model can accurately predict the future position given correct parameters and initial states. However, it is often very complex and almost impossible in reality to build a model including all influencing factors accurately and effectively. Thus physical model is usually developed for simulation system. Learning model can predict future position in a probability way given only the historical motion data. It packages all system internal states and influencing factors into one model. However, the performance depends significantly upon the quality of the data, the learning ability of the model. Hybrid model can combine the advantages of its sub-models and thus is expected to perform better.
\item For collision detection, ship domain is important. Circular or elliptical are usually adopted in open sea because of their simplicity and easy manipulation. Complex shape domains are usually used in restricted sea area (port) in order to make use of limited space. Recently learning domain becomes popular because it can take many influencing factors into consideration and thus make the ship domain self-adaptive.  Most collision risk assessment methods are usually based on constant speed and course at the detection moment. We think that it will be more meaningful to learn a ship's motion features from its large historical AIS data to predict the future path, rather than assuming its motion to be constant.
\item Path planning is an important step for autonomous navigation and collision avoidance. However, it is quite different to planning a ship's path from planning a path for a robot or a land vehicle, due to the unique characteristics of ship motion and sea environment. All most all existing methods works on relatively ideal conditions, such as low traffic, regular or constant moving speed, water current influence and good weather conditions. Therefore, there is still much space to be explored for ship path planning.
\end{itemize}


\ifCLASSOPTIONcaptionsoff
  \newpage
\fi

\bibliographystyle{IEEEtran}
\bibliography{IEEEabrv,refs}

\begin{IEEEbiography}[{\includegraphics[width=1in,height=1.25in,clip,keepaspectratio]{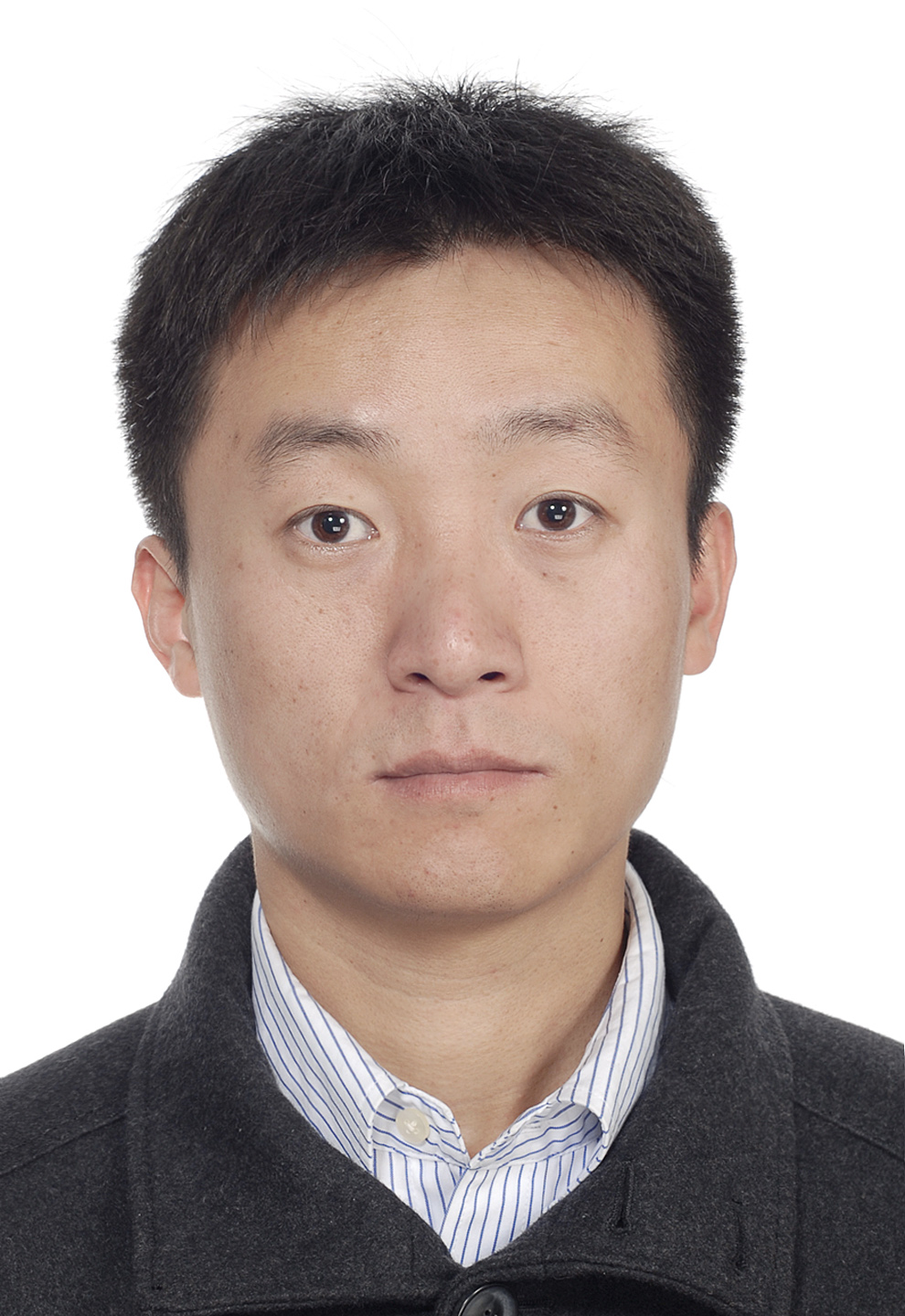}}]{Enmei Tu}
was born in Anhui, China. He received his B.Sc. degree and M.Sc. degree from University of Electronic Science and Technology of China (UESTC) in 2007 and 2010, respectively and PhD degree from the Institute of Image Processing and Pattern Recognition, Shanghai Jiao Tong University, China in 2014. He is now a research fellow in Roll-Royce@NTU Corporate Laboratory at Nanyang Technological University. His research interests are machine learning, computer vision and neural information processing.
\end{IEEEbiography}

\begin{IEEEbiography}[{\includegraphics[width=1in,height=1.25in,clip,keepaspectratio]{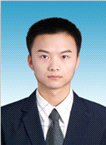}}]{Guanghao Zhang }
received the B.Eng. degree in software engineering from Shandong University and M.Eng. degree in computer science and technology from Beijing Institute of Technology, PR China, in 2012 and 2015, respectively. He is currently a Ph.D. student with School of Electrical and Electronic Engineering, Nanyang Technological University, Singapore. His research interests include neural networks and one-shot learning.
\end{IEEEbiography}

\begin{IEEEbiography}[{\includegraphics[width=1in,height=1.25in,clip,keepaspectratio]{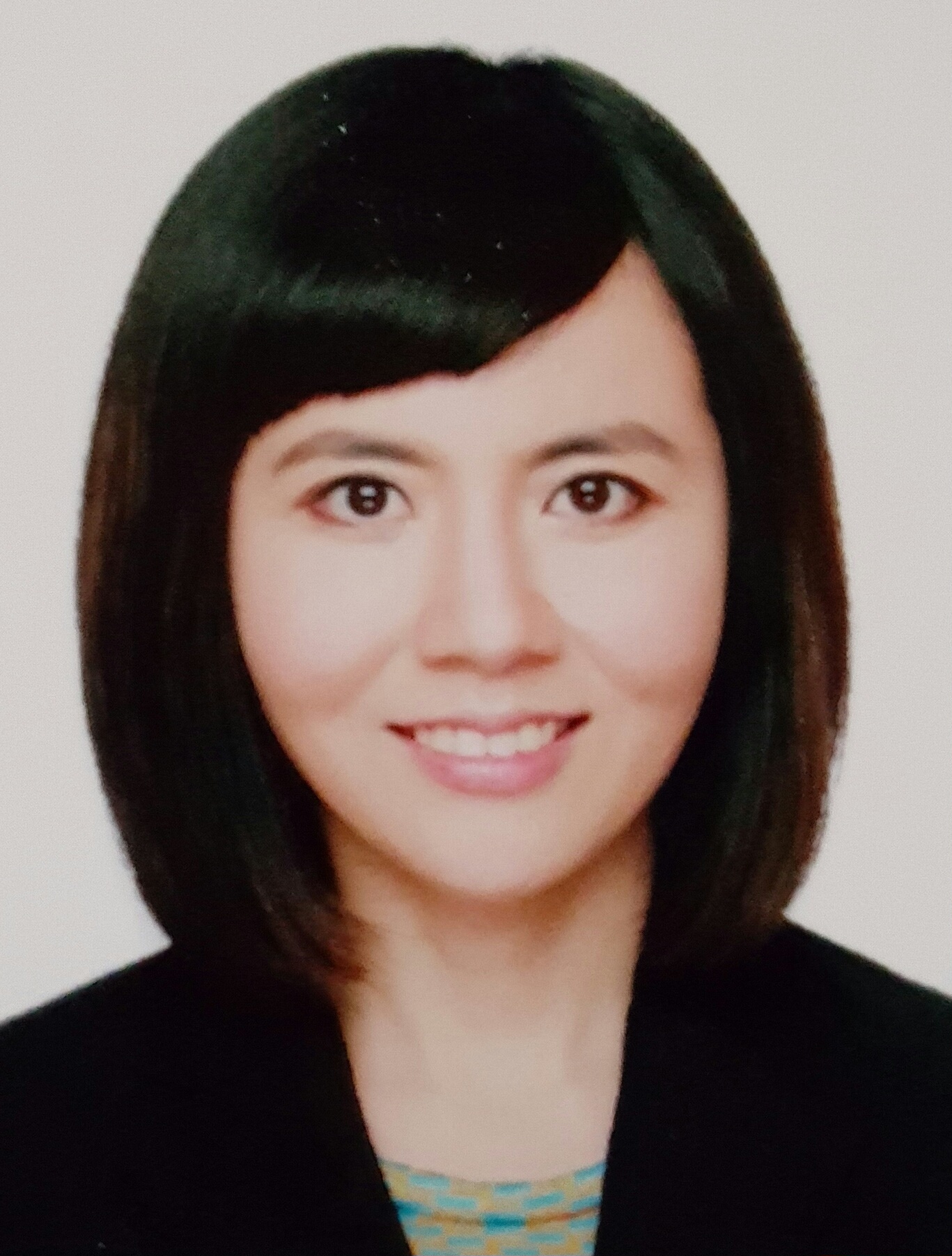}}]{Lily Rachmawati}
received a PhD from the Electrical and Computer Engineering Department, National University of Singapore in 2010 for her work in Multi Objective Evolutionary Algorithms. As part of the computational engineering team in Rolls Royce Singapore, she develops data analytics and optimization techniques for real world problems.
\end{IEEEbiography}

\begin{IEEEbiography}[{\includegraphics[width=1in,height=1.25in,clip,keepaspectratio]{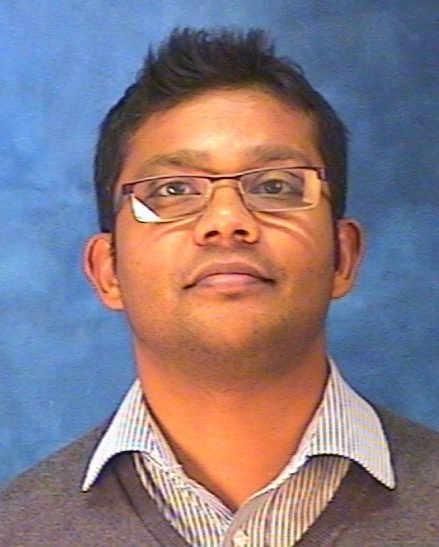}}]{Eshan Rajabally}
 currently manages technology development for the unmanned and autonomous operation of commercial vessels at the Rolls-Royce Strategic Research Centre. With a background in industrially led research and innovation, he has developed a keen interest in the exploitation of computational engineering.
\end{IEEEbiography}

\begin{IEEEbiography}[{\includegraphics[width=1in,height=1.25in,clip,keepaspectratio]{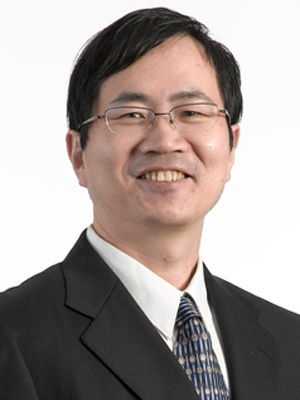}}]{Guang-Bin Huang}
 received the B.Sc  degree in applied mathematics  and M.Eng degree in computer engineer- ing from Northeastern University, PR China, in 1991 and 1994, respectively,  and Ph.D  degree in electrical engineering from Nanyang Technological University, Singapore in 1999.
 From May 2001, he has been working as  an Assistant Professor and Associate Professor (tenured) in the School of Electrical and Electronic Engineering, Nanyang Technological University.  He is  a member of  the Committee
on Membership Development of  IEEE Singapore Chapter.  He serves as session chair,  track chair and
plenary talk chair for  several international conferences and an associate editor of Neurocomputing, and IEEE Transactions on Systems, Man, and Cybernetics – Part B.



\end{IEEEbiography}

\newpage
\onecolumn
\section*{appendices}

\begin{table*}[!h]
\renewcommand{\arraystretch}{1.3}
\caption{Typical AIS message types}
\label{SummaryofAIStypes}
\centering
\begin{tabular}{|l|l|l|}  \hline
\textbf{ID} & \textbf{Type} & \textbf{Description\cite{navcenter}} \\ \hline \hline
1-3 & Position report & (Assigned) Scheduled position report, or response to interrogation\\ \cline{1-3}
4 & Base station report & Position, UTC, date and current slot number of base station \\ \cline{1-3}
5 & Static and voyage related data & Scheduled static and voyage related vessel data report   \\ \cline{1-3}
6-8 & Binary related message & Binary communication \\ \cline{1-3}
10-11 & UTC related message& Request/Response-to UTC/date \\ \cline{1-3}
12-14 & Safety related message & Communication/Acknowledgement/Broadcast safety data\\ \cline{1-3}
15 & Interrogation & Request for special response \\ \cline{1-3}
21 & Aids-to-navigation report & Position and status report for aids-to-navigation \\ \cline{1-3}
27 & Position report for long range applications& Class A and Class B "SO" ship borne mobile equipment outside base station coverage   \\ \cline{1-3}
\end{tabular}
\end{table*}

\begin{table*}[!h]
\renewcommand{\arraystretch}{1.2}
\caption{Position report message}
\label{positionreport}
\centering
\begin{tabular*}{0.93\linewidth}{|p{0.1\linewidth}|p{0.7845\linewidth}|}\cline{1-2}
\textbf{Field} & \textbf{Description\cite{navcenter}}   \\ \hline\hline
MMSI & The vessel's Maritime Mobile Service Identity (MMSI)   \\ \cline{1-2}
Navigational status & 0 = under way using engine, 1 = at anchor, 2 = not under command, 3 = restricted manoeuvre ability, 4 = constrained by her draught, 5 = moored, 6 = aground, 7 = engaged in fishing, 8 = under way sailing, etc.   \\ \cline{1-2}
Rate of turn & Right or left, from 0 to 720 degrees per minute   \\\cline{1-2}
SOG & Knots(0-102.2) Speed over ground in 1/10 knot steps (0-102.2 knots) 1 023 = not available, 1 022 = 102.2 knots or higher  \\ \cline{1-2}
COG & Degrees(0-359). Course over ground in 1/10 = (0-359). 3600 (E10h) = not available = default. 3601-4095 should't be used   \\ \cline{1-2}
Heading & Degrees (0-359) (511 indicates not available = default)   \\ \cline{1-2}
Longitude & Longitude- to 0.0001 minutes   \\ \cline{1-2}
Latitude & Latitude- to 0.0001 minutes   \\ \cline{1-2}
IMO & IMO ship identification number–a seven digit number that remains unchanged \\ \cline{1-2}
Draught & Draught of ship – 0.1 meter to 25.5 meters   \\ \cline{1-2}
Destination & Destination – max. 20 characters   \\ \cline{1-2}
ETA & Estimated time of arrival  \\ \cline{1-2}
Position accuracy & 1 = high (\textless= 10 m) 0 = low (\textgreater 10 m)  \\ \cline{1-2}
Time stamp & UTC time \\\cline{1-2}
\end{tabular*}
\end{table*}

\end{document}